\documentclass[10pt,amsmath,amssymb,aps,pra,twocolumn,longbibliography,floatfix,superscriptaddress,nofootinbib,draft]{revtex4-1}

\usepackage{ifdraft}
\ifdraft{\newcommand{\authnote}[3]{{\color{#3} {\bf  (#1:} #2)}}}{\newcommand{\authnote}[3]{}}

\usepackage[colorinlistoftodos,prependcaption,textsize=small,color=green!30,obeyDraft]{todonotes}

\usepackage[utf8]{inputenc}
\usepackage[T1]{fontenc}

\usepackage{dsfont}
\usepackage{times}
\usepackage{txfonts}
\usepackage[mathcal]{euscript}

\usepackage{upgreek}

\usepackage{mathtools}
\usepackage{mleftright}\mleftright
\usepackage[pagebackref,final,breaklinks,colorlinks,allcolors = blue]{hyperref}
\usepackage{cleveref}
\usepackage{float}

\usepackage{enumitem}

\usepackage{tikz}
\usetikzlibrary{calc,arrows,quotes,angles}

\usepackage{thmtools} 
\usepackage{thm-restate}

\usepackage{multirow}
\usepackage{diagbox}

\floatstyle{boxed}

\newfloat{algorithm}{t}{lop}
\floatname{algorithm}{Algorithm} 

\newfloat{metaalgorithm}{t}{lop}
\floatname{metaalgorithm}{Meta-Algorithm} 

\newtheorem{theorem}{Theorem} 

\newtheorem{lemma}[theorem]{Lemma}

\newtheorem{definition}[theorem]{Definition}

\mathchardef\mhyphen="2D

\newcommand{\Ord}[1]{\mathcal{O}\mleft( #1 \mright)}
\newcommand{\tOrd}[1]{\tilde{\mathcal{O}}\mleft( #1 \mright)}

\newcommand{\ket}[1]{|#1\rangle}
\newcommand{\bra}[1]{\langle#1|}

\newcommand{\ketbra}[2]{|#1\rangle\! \langle #2|}

\newcommand{\Tr}{\mbox{\rm Tr}}

\newcommand{\eps}{\varepsilon}
\renewcommand{\epsilon}{\varepsilon}
\newcommand{\nrm}[1]{\left\lVert#1\right\rVert}
\def\01{\{0,1\}}

\newcommand{\bigO}[1]{\mathcal{O}\left( #1 \right)}
\newcommand{\bOt}[1]{\widetilde{\mathcal O}\left(#1\right)}

\usepackage{graphicx}
\usepackage{dcolumn}
\usepackage{bm}

\usepackage{algorithm}
\usepackage{algpseudocode}

\newcommand{\tPhi}{\widetilde{\Phi}}
\newcommand{\tN}{\widetilde{\mathcal{N}}}
\newcommand{\mN}{\mathcal{N}}
\newcommand{\tE}{\tilde{E}}

\newenvironment{proof}[1][Proof]{\noindent\textbf{#1.} }{\ \rule{0.5em}{0.5em}}

\allowdisplaybreaks

\begin{document}

\title{Quantum algorithm for Petz recovery channels and pretty good measurements}





\author{András Gilyén}
 \email{gilyen@berkeley.edu}
 \affiliation{Institute for Quantum Information and Matter, California Institute of Technology, Pasadena, California 91125, USA}
 \affiliation{Simons Institute for the Theory of Computing, Berkeley, California 94720, USA} 
 \author{Seth Lloyd}
 \email{slloyd@mit.edu}
 \affiliation{Department of Mechanical Engineering and Research Laboratory of Electronics, Massachusetts Institute of Technology, Cambridge, Massachusetts 02139, USA}
 \author{Iman Marvian}
 \email{iman.marvian@duke.edu}
 \affiliation{Department of Physics and Department of Electrical and Computer Engineering,
Duke University, Durham, North Carolina 27708, USA}
 \author{Yihui Quek}
 \email{yquek@stanford.edu}
 \affiliation{Information Systems Laboratory, Stanford University, Stanford, California 94305, USA}
 \author{Mark M. Wilde}
 \email{mwilde@lsu.edu}
 \affiliation{Hearne Institute for Theoretical Physics, Department of Physics and Astronomy, and Center for Computation and Technology, Louisiana State University, Baton Rouge, Louisiana 70803, USA}
 \affiliation{Stanford Institute for Theoretical Physics, Stanford University, Stanford, California 94305, USA}

\date{\today}

\begin{abstract}
The Petz recovery channel plays an important role in quantum information science as an operation that approximately reverses the effect of a quantum channel. The pretty good measurement is a special case of the Petz recovery channel, and it allows for near-optimal state discrimination. A hurdle to the experimental realization of these vaunted theoretical tools is the lack of a systematic and efficient method to implement them. This paper sets out to rectify this lack: using the recently developed tools of quantum singular value transformation and oblivious amplitude amplification, we provide a quantum algorithm to implement the Petz recovery channel when given the ability to perform the channel that one wishes to reverse. Moreover, we prove that, in some sense, our quantum algorithm's usage of the channel implementation cannot be improved by more than a quadratic factor. Our quantum algorithm also provides a procedure to perform pretty good measurements when given multiple copies of the states that one is trying to distinguish. 
\end{abstract}

\maketitle


{\em Introduction}---Pretty good measurements \cite{Belavkin75,Belavkin75a,HolPGM78,Hausladen93bach,hausladen94} and Petz recovery channels \cite{petz1986sufficientSubalgebrasAndRelativeEntropy,petz1988sufficiencyOfChannels,OP93,barnum2000reversingQuantumDynamics,hayden2003structureStatesStrongSubaddivityWEquality} are workhorses of quantum information theory: they are used ubiquitously to prove
basic results in quantum communication and measurement \cite{wilde2017QIT}. Although important for attaining quantum channel capacities \cite{hausladen96,Hol98,SW97,BG14,beigi2015decodingQITviaPetz} and performing
state discrimination \cite{Belavkin75,Belavkin75a,HolPGM78,Hausladen93bach,barnum2000reversingQuantumDynamics}, these useful theoretical constructions are less common in experiment, for the simple reason that 
there has not been a systematic method for performing them efficiently in practice. Our goal here is to fill this gap.

The Petz recovery channel was introduced in the context of quantum sufficiency  in \cite{petz1986sufficientSubalgebrasAndRelativeEntropy,petz1988sufficiencyOfChannels} and later rediscovered in \cite{barnum2000reversingQuantumDynamics} in the context of quantum error correction.
It can be understood as a critical part of a quantum version of the Bayes theorem \cite[Section~IV]{leifer2013TowardsQuantThViaBayesianInference}. To review it, let us begin with the classical case.  A classical channel with input system $X$ and output system $Y$ over the alphabets $\mathcal{X},\mathcal{Y}$ is a conditional probability distribution $\{p_{Y|X}(y|x)\}_{x\in \mathcal{X},y\in \mathcal{Y}}$. We consider a probability distribution $p_X(x)$ over the alphabet $\mathcal{X}$ as the input to the channel. It then follows from the Bayes theorem that
$p_X(x) p_{Y|X}(y|x) = p_Y(y) p_{X|Y}(x|y)$, where $p_Y(y) = \sum_x p_X(x) p_{Y|X}(y|x)$. Hence, for all  $x\in \mathcal{X}, y\in\mathcal{Y}$, we define the ``reversal channel'' via the formula
\begin{equation}
    p_{X|Y}(x|y) = \frac{p_X(x) p_{Y|X}(y|x)}{\sum_x p_X(x) p_{Y|X}(y|x)}.
\end{equation}
This channel acts on the output system $Y$. If the particular distribution $p_Y(y)$ defined above is ``sent in'' through this channel, then the input $p_X(x)$ is recovered perfectly: $p_X(x) = \sum_y p_{X|Y}(x|y) p_Y(y)$. The computation of the reversal channel $p_{X|Y}(x|y)$ requires a specification of the input probability distribution $p_X(x)$ and the forward channel $p_{Y|X}(y|x)$. The Petz recovery channel is a quantum generalization of the reversal channel above: it is a function of a quantum channel $\mathcal{N}$ and an input state $\sigma$ to the channel, with the former generalizing $p_{Y|X}(y|x)$ and the latter $p_{X}(x)$. We discuss it in more detail in what follows.

The Petz recovery channel appears often in quantum information as a proof tool, showing that near-optimal recovery from undesired quantum operations is possible. Ref.~\cite{barnum2000reversingQuantumDynamics} demonstrated how this recovery channel can be an effective means for reversing the effects of noise. Thereafter, \cite{Ng_2010} showed that the Petz recovery channel (therein called ``transpose channel'') is a universal recovery operation for approximate quantum error correction, which performs comparably to the best possible one in terms of worst-case fidelity (see also \cite{tyson10,MN2012}).
The Petz recovery channel also goes by the name ``pretty good recovery,'' as used in \cite{DFW15,PhysRevA.96.042328}, due to the result of \cite{barnum2000reversingQuantumDynamics}.
Yet another application comes from the field of quantum communication: \cite{beigi2015decodingQITviaPetz} showed explicitly how to use the Petz recovery channel in a decoder to achieve the coherent information rate of quantum communication. It has also found use in developing physically meaningful refinements of quantum entropy inequalities \cite{wilde2015recoverabilityInQITheor,Sutter15,Junge15,Carlen_2020,faulkner2020}.
See \cite{Lami_2018,swingle2019,CHPSSW19,jia2020,PSSY19} for further uses.

As an application of our results, our quantum algorithm can be used to implement the pretty good measurement (PGM) \cite{Belavkin75,Belavkin75a,HolPGM78,Hausladen93bach,hausladen94}. This measurement was used in \cite{hausladen96,BG14} as part of a coding scheme to approach the Holevo information rate for classical communication over a quantum channel. It has also been instrumental in proving bounds for quantum algorithms. Ref.~\cite{bacon2005optimalMeasurementDiahedralHSP} showed that the PGM
is an optimal measurement for solving the dihedral hidden subgroup problem and that it is helpful in proving a lower bound on the sample complexity of this problem. Similar techniques have been used for quantum probably-approximately-correct learning \cite{arunachalam2017OptQSampCoplLearn}.
Ref.~\cite{lloyd2020QPolarDecAlg}  showed how to implement the PGM for pure states, while our algorithm for Petz recovery channels is capable of performing the PGM in the general case.


We now begin the technical part of our paper, starting with an explicit description of the Petz recovery channel and the resources that we work with for its implementation.

{\em Petz recovery channel}---The Petz recovery channel is a function of a quantum state $\sigma_{A}$ on a system $A$ and a quantum channel $\mathcal{N}_{A\rightarrow B}$ taking system $A$ to a system $B$. It is given explicitly as
follows \cite{hayden2003structureStatesStrongSubaddivityWEquality}:%
\begin{equation}
\mathcal{P}_{B\rightarrow A}^{\sigma,\mathcal{N}}(\omega_{B})
\coloneqq 
\sigma_A
^{1/2} \mathcal{N}^{\dag}\left(  
\mathcal{N}(\sigma_A)^{-1/2}\omega_{B}
\mathcal{N}(\sigma_A)^{-1/2}\right)  \sigma_A^{1/2}, \label{eq:petz-map}%
\end{equation}
where $\mathcal{N}^{\dag}$ is the Hilbert--Schmidt adjoint \cite{wilde2017QIT} of the channel $\mathcal{N}$  and we have omitted the system labels of $\mathcal{N}_{A\rightarrow B}$ for brevity.

It is a composition of three completely positive (CP)\ maps:%
\begin{align}
(\cdot)  &  \rightarrow\left[  \mathcal{N}(\sigma
_{A})\right]  ^{-1/2}(\cdot)\left[  \mathcal{N}(\sigma
_{A})\right]  ^{-1/2},\label{eq:cp-map-1}\\
(\cdot)  &  \rightarrow  \mathcal{N}^{\dag
}(\cdot),\label{eq:cp-map-2}\\
(\cdot)  &  \rightarrow\sigma_{A}^{1/2}(\cdot)\sigma_{A}^{1/2}.
\label{eq:cp-map-3}%
\end{align}
None of these maps are trace preserving individually, but overall the map in \eqref{eq:petz-map} is trace preserving on the support of the state~$\mathcal{N}(\sigma_{A})$ \cite{wilde2015recoverabilityInQITheor}. We note here that the main idea behind our algorithm is to implement the Petz recovery channel as a composition of the three maps given in \eqref{eq:cp-map-1}--\eqref{eq:cp-map-3}, while taking into account the fact that the overall map in \eqref{eq:petz-map} is trace-preserving in order to implement it deterministically with some desired accuracy. 

{\em Block-encoding}---The Petz recovery channel depends on the state $\sigma_{A}$, and so our algorithm needs some form of access to it. In order to cover a wide range of scenarios, we employ the block-encoding formalism, which generalizes the most common input models for matrices used in quantum algorithms \cite{low2016HamSimQubitization,gilyen2018QSingValTransf}.

Let $\lVert \cdot \rVert$ denote the spectral norm of a matrix (also known as the Schatten $\infty$-norm). 
For a complex matrix $A$ and $\alpha \geq \lVert A \rVert$, the matrix $A/\alpha$ can be represented as the upper-left block of a unitary matrix:
\begin{equation}\label{eq:block-encoding}
U=\left[\begin{array}{cc}A / \alpha & \cdot \\ \cdot & \cdot\end{array}\right] \quad \Longleftrightarrow \quad A=\alpha(\langle 0| \otimes I) U(|0\rangle \otimes I).
\end{equation}
The unitary matrix $U$ is said to be a {\em block-encoding} of $A$. Henceforth, we do not write identity operators explicitly, but we instead include system subscripts as a guide. If the linear map $A/\alpha$ acts on $a$ qubits, then the unitary $U$ can be thought of as a probabilistic implementation of this map: given an $a$-qubit input state $|\psi\rangle,$ applying the unitary $U$ to the state $|0\rangle|\psi\rangle,$ measuring the first system, and post-selecting on the $|0\rangle$ outcome, the second system contains a state proportional to $A \ket{\psi}/\alpha$. 

This generalizes the two most relevant input models in our case. If we are given copies of the quantum state $\sigma_{A}$, then we can implement an (approximate) block-encoding of $\sigma_{A}$ by using density matrix exponentiation~\cite{lloyd2013QPrincipalCompAnal,kimmel2016hamiltonian} and ``taking the logarithm'' of the time evolution~\cite{gilyen2018QSingValTransf}. 
Alternatively, if we have access to a quantum circuit $U_{RA}^{\sigma}$ that prepares a purification 
$|\psi^{\sigma}\rangle_{RA} \coloneqq U_{RA}^{\sigma}|0\rangle_{R}|0\rangle_{A}$
of $\sigma_{A}$, such that
 $\operatorname{Tr}_{R}[\ketbra{\psi^{\sigma}}{\psi^{\sigma}}_{RA}%
]=\sigma_{A},$
then we can directly implement an exact block-encoding of~$\sigma_A$ with only two uses of $U_{RA}^{\sigma}$ as follows \cite{low2016HamSimQubitization,gilyen2018QSingValTransf}: 
\begin{equation}\label{eq:block_encoding_density_op}
V_{RAA^{\prime}}^{\sigma}\coloneqq(U_{RA}^{\sigma})^{\dag}\left(  I_{R}\otimes
\operatorname{SWAP}_{AA^{\prime}}\right)  U_{RA}^{\sigma}=%
\begin{bmatrix}
\sigma_{A} & \cdot\\
\cdot & \cdot
\end{bmatrix}
\, ,
\end{equation}
where system $A'$ is isomorphic to system $A$.

{\em Assumptions}---The resources that we use for implementing the Petz recovery channel are as follows: 
1) Quantum circuits $U^{\sigma_A}$ and $U^{\mathcal{N}(\sigma_A)}$ that are (approximate)  block-encodings of $\sigma_{A}$ and $\mathcal{N}(\sigma_A)$, respectively, and
2) a quantum circuit $U_{E^{\prime}A\rightarrow EB}^{\mathcal{N}}$ that implements the channel $\mathcal{N}$, in the sense that 
$U_{E^{\prime}A\rightarrow EB}^{\mathcal{N}}|0\rangle_{E^{\prime}%
} =: V_{A\rightarrow EB}^{\mathcal{N}},$
where $V_{A\rightarrow EB}^{\mathcal{N}}$ is an isometric extension of
$\mathcal{N}$ satisfying $\operatorname{Tr}_{E}[V_{A\rightarrow EB}^{\mathcal{N}}(\omega_{A})(V_{A\rightarrow EB}^{\mathcal{N}})^{\dag}]=\mathcal{N}(\omega_{A}),$ for every input density operator $\omega_{A}$.

We note that, given an efficient description of the channel~$\mathcal{N}$ in terms of its Kraus operators, the unitary $U_{E^{\prime}A\rightarrow EB}^{\mathcal{N}}$ can be efficiently implemented on a quantum computer \cite{cleve2017}. Also, given copies or ``purified access'' to $\sigma_{A}$, we can achieve the corresponding access to $\mathcal{N}(\sigma_A)$ after applying $U_{E^{\prime}A\rightarrow EB}^{\mathcal{N}}$, which then results in an efficient block-encoding for $\mathcal{N}(\sigma_A)$.

{\em Rewriting the Petz recovery channel}---Eq.~\eqref{eq:cp-map-2} calls for the application of the adjoint $ \mathcal{N}^{\dag}$ of the channel $ \mathcal{N}$. We now explain how this can be accomplished using $U_{E^{\prime}A\rightarrow EB}^{\mathcal{N}}$. The action of the adjoint on an arbitrary operator $\omega_B$ is given by
$\mathcal{N}^{\dag}(\omega_{B})=\langle
0|_{E^{\prime}}U^{\mathcal{N}\, \dag}\left(
I_{E} \otimes \omega_{B} \right)  U^{\mathcal{N}%
}|0\rangle_{E^{\prime}}$ \cite{wilde2017QIT}.
Let $\Gamma_{E \tilde{E} } \coloneqq \ketbra{\Gamma}{\Gamma}_{E \tilde{E}}$ denote an operator proportional to the maximally entangled state on $E$ and a reference system $\tilde{E}$, where $|\Gamma\rangle_{E \tilde{E}} \coloneqq  \sum_{i=0}^{d_{E}-1}|i\rangle_{E}|i\rangle_{\tilde{E}}$ and $d_E$ is the dimension of system $E$.
Then extending the identity operator with $\Gamma_{E \tilde{E} }$, we rewrite the previous identity as
\begin{equation}
\mathcal{N}^{\dag}(\omega_{B}) = \operatorname{Tr}_{\tilde{E}}[\langle0|_{E^{\prime}}(U_{E^{\prime
}A\rightarrow EB}^{\mathcal{N}})^{\dag}\left(  \Gamma
_{E \tilde{E} } \otimes \omega_{B}\right)  U_{E^{\prime}A\rightarrow EB}^{\mathcal{N}}%
|0\rangle_{E^{\prime}}].\label{eq:extension-adjoint-2}%
\end{equation}

Now the interpretation of the adjoint map as a probabilistic quantum operation is clear: the adjoint map $\mathcal{N}^{\dag}$ acting on the operator $\omega_B$ can be applied by tensoring in the maximally entangled state $\Gamma_{E \tilde{E}} / d_E$, performing the inverse of the unitary~$U^{\mathcal{N}}$, measuring the system $E^{\prime}$, accepting if the all-zeros outcome occurs, and finally, ignoring the system~$\tilde{E}$ (which corresponds to tracing it out). 

Thus, our plan is to implement  the linear extension of the adjoint map, as given in \eqref{eq:extension-adjoint-2}. Sandwiching this between the other two maps in \eqref{eq:cp-map-1} and \eqref{eq:cp-map-3} comprising the Petz recovery channel, we obtain the following isometric extension of the Petz recovery channel:%
\begin{equation}
V_{B\rightarrow \tilde{E} A}^{\mathcal{P}}\coloneqq  
(\bra{0}_{E^{\prime}}\otimes I_{\tilde{E}}\otimes\sigma_{A}^{\frac{1}{2}}
)(U_{E^{\prime}A\rightarrow EB}^{\mathcal{N}})^{\!\dag}
(\ket{\Gamma}_{\!E \tilde{E} }\otimes \left[\mathcal{N}(\sigma_{A})\right]^{-\frac{1}{2}}).\label{eq:iso-ext-petz-map}
\end{equation}
Tracing over $\tilde{E}$ then implements the Petz recovery channel $\mathcal{P}_{B\rightarrow A}^{\sigma,\mathcal{N}}(\omega_{B})$. Note that in the rewriting above, the implementation of the adjoint map discussed in the preceding paragraph is no longer contiguous. It proceeds in two phases: the application of the unitary $(U_{E^{\prime}A\rightarrow EB}^{\mathcal{N}})^{\dag}$ before multiplication by $\sigma_{A}^{1/2}$ (which applies \eqref{eq:cp-map-3}); and the measurement and post-selection after that step.

{\em Quantum singular value transformation}---Our implementation is based on quantum singular value transformation (QSVT) \cite{gilyen2018QSingValTransf}. QSVT transforms the singular values of a block-encoded matrix and thus provides an efficient means of quantum matrix arithmetic. Often we need to rely on approximations, and so when doing so, we keep track of the error/precision $\delta$, as well as the sub-normalization factor $\alpha$: we say that $U$ is an $(\alpha,\delta)$-block-encoding of $A$ if $\left\|A-\alpha(\langle 0| \otimes I ) U(|0\rangle \otimes I) \right\| \leq \delta.$

In what follows, we manipulate block-encodings $U^\rho$ of density operators $\rho$. The power of QSVT is that it allows for transforming $U^\rho$ to a block-encoding of $\tilde{f}(\rho)$, where $\tilde{f}$ is a function applied to the singular values of its argument. More precisely, $\tilde{f}$ denotes a polynomial approximation of some function $f$; in view of the maps given in \eqref{eq:cp-map-1} and \eqref{eq:cp-map-3} above, the particular functions of interest here are $f_1(x) \coloneqq x^{-1/2}$ and $f_2(x) \coloneqq x^{1/2}$. 

The complexity of realizing the transformed block-encoding unitary $U^{\tilde{f}(\rho)}$ is stated in terms of the number of uses of $U^\rho$ (which dominates the overall gate complexity), and it depends on the parameters of the functional approximation~$\tilde{f}$. 
For a function $f$, let $\left\Vert f(x)\right\Vert _{\mathcal{I}}\coloneqq\sup_{x\in\mathcal{I}}\left\vert f(x)\right\vert$. 
Using techniques from \cite{gilyen2018QSingValTransfThesis}, for the two functions above, one can find polynomial approximations $\tilde{f}_1$, $\tilde{f}_2$ such that $\frac{\theta^{1/2}}{2}\left\Vert \tilde{f}_1(x)-x^{-1/2}%
\right\Vert _{\left[  \theta,1\right]  }\leq\delta,$ and $\frac{1}{2}\left\Vert \tilde{f}_2(x)-x^{1/2}\right\Vert _{\left[
\theta,1\right]  }\leq\delta$ for $\theta
,\delta\in(0,1/2]$. If $\rho$ has minimum singular value $\lambda_{\min}$, then it suffices to set $\theta \leq \lambda_{\min}$. Since $1/\lambda_{\min}$ behaves like a ``condition number'' for $\rho$, being proportional to the difficulty of transforming $\rho$, we denote it with the symbol $\upkappa$ and employ this notation later.  Indeed, using the functional approximations from~\cite{gilyen2018QSingValTransfThesis,gilyen2018QSingValTransfThesis}, QSVT achieves the desired transformations up to the errors indicated above, with $\bigO{  \frac{1}{\theta}\log\frac{1}{\delta}}$ uses~of~$U^{\rho}$. 

{\em The quantum algorithm}---We implement the isometric extension of the Petz recovery channel given in \eqref{eq:iso-ext-petz-map}. This consists of applying the maps in \eqref{eq:cp-map-1}, \eqref{eq:cp-map-2}, and \eqref{eq:cp-map-3} sequentially, with the first and third steps employing QSVT. Eq.~\eqref{eq:iso-ext-petz-map} also has a measurement component as the final step, arising from the implementation of the map in \eqref{eq:cp-map-2}. By exploiting the trace-preserving property of the Petz recovery channel, we amplify the probability of success of this measurement (i.e., the projection onto $\ket{0}_{E^{\prime}}$) using oblivious amplitude amplification~\cite{berry2013ExpPrecHamSimSTOC}, which is a special case of QSVT~\cite{gilyen2018QSingValTransf}. Overall, the implementation is precise up to $\eps$ error in diamond distance \cite{Kitaev1997} (see \cite{wilde2017QIT} for a definition of diamond distance). Theorem~\ref{thm:petz} below states the guarantees of this technique.

\begin{theorem}\label{thm:petz}
Let $N_{\sigma}$, $N_{\mathcal{N}(\sigma)}$ and $N_{\mathcal{N}}$
denote the number of elementary quantum gates needed to realize the unitaries
$U^{\sigma_A}$, $U^{\mathcal{N}(\sigma_A)}$, and $U_{E^{\prime} A \rightarrow
E B}^{\mathcal{N}}$, respectively (noting that without loss of generality $N_{\mathcal{N}(\sigma)} \leq N_{\sigma}+2N_{\mathcal{N}}$). 
Let $\upkappa_{\sigma}$ denote an upper bound on the reciprocal of the minimum non-zero eigenvalue of $\sigma_A$, and correspondingly, let $\upkappa_{\mathcal{N}(\sigma)}$ denote the same for $\mathcal{N}(\sigma_A)$.
There exists a quantum algorithm realizing the channel $\mathcal{\tilde{P}}_{B\rightarrow A}^{\sigma_A,\mathcal{N}}$, which is an approximate implementation of the ideal Petz recovery channel in \eqref{eq:petz-map}, in the sense that
\begin{equation}
\left\Vert \mathcal{\tilde{P}}_{B\rightarrow A}^{\sigma_A,\mathcal{N}}-
\mathcal{P}_{B\rightarrow A}^{\sigma_A,\mathcal{N}%
}
\right\Vert
_{\diamond}\leq\eps ,\label{eq:final-diamond-norm-bnd-petz}%
\end{equation} with gate complexity (up to poly-logarithmic factors)
\begin{equation}
\label{eq:performance-petz-map}
\bOt{\sqrt{d_{E}\upkappa_{\mathcal{N}(\sigma)}}
\left(
\upkappa_{\mathcal{N}(\sigma)} N_{\mathcal{N}(\sigma)}
\!+\! N_{\mathcal{N}} \!+\! N_{\sigma}\min\left(\upkappa_{\sigma}, d_E \upkappa_{\mathcal{N}(\sigma)}/{\eps^2}\right)\right)}.
\end{equation}
In \eqref{eq:performance-petz-map}, $d_E$ is the dimension of the system $E$, which is not smaller than the Kraus rank of the channel $\mathcal{N}(\cdot)$.
\end{theorem}

In the Supplementary Material \footnote{The Supplementary Material also includes Ref.~\cite{apeldoorn2018ImprovedQSDPSolving}.}, we provide a  modified algorithm that substitutes the dependence on $d_E$ in \eqref{eq:performance-petz-map} with the rank of the state $\widetilde{\mathcal{N}}(\sigma)$, where $\widetilde{\mathcal{N}}$ is a channel complementary to $\mathcal{N}$ \cite{wilde2017QIT}. For certain choices of $\mathcal{N}$ and $\sigma$, this provides a dramatic reduction in the running time.

We now break the algorithm down into its four steps and analyze each step individually (assuming without loss of generality that $\eps=\bigO{1}$). We indicate the steps using the numbers $\bm{(1)}$-$\bm{(4)}$.

$\bm{(1)}$ To simulate the first step of the Petz recovery channel, as described by \eqref{eq:cp-map-1}, we transform the block-encoding of $ \mathcal{N}(\sigma_{A})$ to a $\left(2 \sqrt{\upkappa_{\mathcal{N}(\sigma)}},\frac{\bigO{\eps}}{\sqrt{d_E}} \right)$-block-encoding $U_{R^{\prime}B}^{\tilde{f}_1(\mathcal{N}(\sigma_A))}$ of $\left[  \mathcal{N}(\sigma_{A})\right]^{-1/2}$ using QSVT, which has gate complexity $\bigO{  \upkappa_{\mathcal{N}(\sigma)}N_{\mathcal{N}(\sigma)}\log\frac{d_E}{\varepsilon}}$. Then the following error bound holds
\begin{equation}\label{eq:1sterrorbd}
\left\Vert \tilde{f}_1(\mathcal{N}(\sigma_{A}))-(\mathcal{N}(\sigma_{A}))^{-1/2}\right\Vert\leq \bigO{\eps}/\sqrt{d_E} ,
\end{equation}
which suffices for our purposes, as shown later.

$\bm{(2)}$ Let $\tilde{E}$ be a system with dimension equal to that of $E$. The second step of the algorithm is simply to prepare the maximally entangled state $|\Phi\rangle_{E \tilde{E}}\coloneqq |\Gamma\rangle
_{E \tilde{E}} / \sqrt{d_E}$ alongside the state prepared above, and then apply the unitary $(U_{AE^{\prime
}\rightarrow BE}^{\mathcal{N}})^{\dag}$. Note that $|\Phi\rangle
_{E \tilde{E}}$ is a normalized quantum state, introducing an additional factor of $\frac{1}{d_{E}}$ in the output density operator, which resurfaces in the subnormalization factor of the overall unitary (see~\eqref{eq:W-tilde-as-block-enc}). The maximally entangled state $|\Phi\rangle_{E \tilde{E}}$ is prepared by means of a unitary $U^{\Phi}_{E \tilde{E}}$ acting on the state $\ket{0}_{E \tilde{E}}$, so that $|\Phi\rangle
_{E \tilde{E}} \coloneqq U^{\Phi}_{E \tilde{E}}\ket{0}_{E \tilde{E}}$. Note that the unitary $U^{\Phi}_{E \tilde{E}}$ is easy to implement. For example, if systems $E$ and $\tilde{E}$ consist of qubits, one can apply Hadamard gates on the qubits of $E$ and CNOT gates between pairs of qubits of $E$ and $\tilde{E}$. In this step, we have described the first half of the procedure for implementing a linear extension of~\eqref{eq:cp-map-2}; the final part, which consists of measurement and post-selection, is deferred to the fourth step. 

$\bm{(3)}$ The third step of the algorithm is to apply an approximation of the map in \eqref{eq:cp-map-3} that conjugates the state by $\sigma_A^{1/2}$. Analogous to the first step, we transform the block-encoding of $\sigma_{A}$ to a $\left(2,\frac{\bigO{\eps}}{\sqrt{d_E \upkappa_{\mathcal{N}(\sigma)}}}\right)$-block-encoding $U^{\tilde{f}_2(\sigma_{A})}_{R^{\prime\prime}A}$ of $\tilde{f}_2(\sigma_{A})$ using QSVT, which has gate complexity $\bigO{  \upkappa_{\sigma} N_{\sigma}\log \left(\frac{d_E \upkappa_{\mathcal{N}(\sigma)}}{\varepsilon}\right)}$. Then the following error bound holds
\begin{equation}\label{eq:2nderrorbound}
\left\Vert \tilde{f}_2(\sigma_{A})-\sigma_{A}^{1/2}\right\Vert
\leq \bigO{\eps}/\sqrt{d_E \upkappa_{\mathcal{N}(\sigma)}}.
\end{equation}

We can now apply the unitary $U^{\tilde{f}_2(\sigma_{A})}_{R^{\prime\prime}A}$ to the output of Step~2. In detail, letting $\rho_{A}$ denote the output state of Step~2, we tensor in the state $\ketbra{0}{0}_{R^{\prime\prime}}$ to the input state $\rho_{A}$ and perform the unitary $U^{\tilde{f}_2(\sigma_{A})}_{R^{\prime\prime}A}$.

Let us summarize the algorithm up to this point. We have described the addition of auxiliary systems as happening separately in each step. However, we are free to tensor them in to the input state $\omega_{B}$ at the start, enlarging the input state to
$\ketbra{0}{0}_{R^{\prime\prime}}\otimes\ketbra{0}{0}
_{E \tilde{E}}\otimes\ketbra{0}{0}_{R^{\prime}}\otimes\omega_B$. Then to this state, we apply the following product of unitaries:
\begin{equation}
 \tilde{W}\coloneqq U^{\tilde{f}_2(\sigma_{A})}_{R^{\prime\prime}A} 
 \left(  U_{E^{\prime}A\rightarrow EB}^{\mathcal{N}}\right)^{\dag}
 \left( U^{\Phi}_{E \tilde{E}} \otimes
 U_{R^{\prime}B}^{\tilde{f}_1(\mathcal{N}(\sigma_{A}))} \right) ,
\end{equation}
where $U^{\tilde{f}_2(\sigma_{A})}_{R^{\prime\prime}A}$ and $ U_{R^{\prime}B}^{\tilde{f}_1(\mathcal{N}(\sigma_{A}))} $ are implemented using QSVT. The unitary $\tilde{W}$ approximates the isometric extension in \eqref{eq:iso-ext-petz-map} and can be represented as the following block-encoding:
\begin{equation} 
\tilde{W}=
\begin{bmatrix}
\frac{1}{4}\sqrt{\frac{1}{d_{E}\upkappa_{\mathcal{N}(\sigma)}}}\tilde{V}_{B\rightarrow  \tilde{E} A}^{\mathcal{P}} & \cdot \\
\cdot & \cdot
\end{bmatrix}
,
\label{eq:W-tilde-as-block-enc}
\end{equation}
where the linear operator $\tilde{V}_{B\rightarrow  \tilde{E} A}^{\mathcal{P}}$ is an approximate isometric extension of the Petz recovery channel and is defined through its action on a ket $|\psi\rangle_{B}$ as
\begin{equation}
    \tilde{V}_{B\rightarrow  \tilde{E} A}^{\mathcal{P}}|\psi\rangle_{B}
    \coloneqq 
\tilde{f}_2(\sigma_{A})\left(V_{A\rightarrow EB}^{\mathcal{N}}\right)^{\dag}\tilde{f}_1(\mathcal{N}(\sigma_{A}))   \ket{\Gamma}_{E \tilde{E}} \ket{\psi}_{B}.
\label{eq:approx-iso-petz}
\end{equation}

After applying $\tilde{W}$ to the enlarged input state, we would like to measure the $R^{\prime\prime}E^{\prime}R^{\prime}$ systems and obtain the all-zeros state as the outcome (which corresponds to the top-left block of $\tilde{W}$). Receiving this outcome signals the successful implementation of the desired map $\tilde{V}_{B\rightarrow  \tilde{E} A}^{\mathcal{P}}$,
up to a sub-normalization factor of $4\sqrt{\upkappa_{\mathcal{N}(\sigma)}d_{E}}$. To compare this to the ideal isometric extension in \eqref{eq:iso-ext-petz-map}, we should account for the accumulated errors due to the approximate implementations of $ \mathcal{N}(\sigma_{A}) ^{-1/2}$ and $\sigma_A^{1/2}$ in~$\tilde{W}$. It follows that 
\begin{equation}
\nrm{\tilde{V}_{B\rightarrow  \tilde{E} A}^{\mathcal{P}} - V_{B\rightarrow  \tilde{E} A}^{\mathcal{P}}  } \leq \bigO{\eps} , \label{eq:error_bound}
\end{equation}
where $\tilde{V}_{B\rightarrow  \tilde{E} A}^{\mathcal{P}}$ is defined in \eqref{eq:approx-iso-petz} and $V_{B\rightarrow  \tilde{E} A}^{\mathcal{P}}$  in \eqref{eq:iso-ext-petz-map}. To see this, observe that the left-hand side of \eqref{eq:error_bound} can be bounded from above by the following quantity:
\begin{multline}
\left\Vert \sigma_{A}^{1/2}-\tilde{f}_{2}(\sigma_{A})\right\Vert\ 
\left\Vert \left(  V^{\mathcal{N}}_{A\rightarrow EB}\right)  ^{\dag}\mathcal{N}(\sigma
_{A})^{-1/2}|\Gamma\rangle_{E\tilde{E}}\right\Vert \ +  \\
\left\Vert \tilde{f}_{2}(\sigma_{A})\left(  V^{\mathcal{N}}_{A\rightarrow
EB}\right)  ^{\dag}\right\Vert\ 
\left\Vert |\Gamma\rangle_{E\tilde{E}}\right\Vert\  \left\Vert
\mathcal{N}(\sigma_{A})^{-1/2}-\tilde{f}_{1}(\mathcal{N}(\sigma_{A}%
))\right\Vert,
\label{eq:err-upper-bnd-bnd}
\end{multline}
which follows from applying the triangle inequality and submultiplicativity of the spectral norm.
Noting that $|\Gamma\rangle_{E\tilde{E}}$ is the unnormalized maximally-entangled vector, we further bound the following terms:
\begin{align}
\left\Vert \left(  V^{\mathcal{N}}_{A\rightarrow EB}\right)  ^{\dag}\mathcal{N}(\sigma
_{A})^{-1/2}|\Gamma\rangle_{E\tilde{E}}\right\Vert
& \leq \sqrt{d_E \upkappa_{\mathcal{N}(\sigma_{A})}} , \label{eq:triangle-1}
\\
\left\Vert \tilde{f}_{2}(\sigma_{A})\left(  V^{\mathcal{N}}_{A\rightarrow
EB}\right)  ^{\dag}\right\Vert\ 
\left\Vert |\Gamma\rangle_{E\tilde{E}}\right\Vert
&
\leq 2\sqrt{d_E}.
\label{eq:triangle-2}
\end{align}
The second bound follows because $\tilde{f}_2(\sigma_{A})$ is a block-encoding with norm at most $2$. Putting \eqref{eq:err-upper-bnd-bnd}--\eqref{eq:triangle-2} together with the bounds in \eqref{eq:1sterrorbd} and \eqref{eq:2nderrorbound}, we conclude an overall error between $V^{\mathcal{P}}$ and $\tilde{V}^{\mathcal{P}}$ of $\bigO{\varepsilon}$.

$\bm{(4)}$ Finally, we move on to the last step, which is a measurement of the $R^{\prime\prime}E^{\prime}R^{\prime}$ systems. Eq.~\eqref{eq:W-tilde-as-block-enc} makes it clear that the probability $p_{\rm success}$ of measuring the all-zeros state, at this point, is approximately $\frac{1}{16d_{E}\upkappa_{\mathcal{N}(\sigma)}}$.  We would like to amplify this probability, and so we use oblivious amplitude amplification to implement an approximate projection onto this state. This too can be achieved using QSVT techniques~\cite{gilyen2018QSingValTransf} and requires a number of repetitions of $\tilde{W}$ that scales as $\bigO{1/\sqrt{p_{\rm success}}}$, which in this case is $N_{\rm rep} \coloneqq \bigO{\sqrt{d_{E}{\upkappa_{\mathcal{N}(\sigma)}}}}$.
After applying (robust) oblivious amplitude amplification~\cite[Theorem 28]{gilyen2018QSingValTransfArXiv}, we obtain a unitary that is a $(1,\bigO{\eps})$-block-encoding of the isometric extension $V_{B\rightarrow \tilde{E} A }^{\mathcal{P}}$, providing an $\bigO{\eps}$-approximate implementation of the Petz recovery channel.

The complexity of our algorithm is given by $N_{\rm rep}$ times the complexity of implementing $\tilde{W}$. As we discussed previously, the cost of implementing the first step in $\tilde{W}$ is $\bigO{  \upkappa_{\mathcal{N}(\sigma)}N_{\mathcal{N}(\sigma)}\log\frac{d_E }{\varepsilon}}$. The complexity of implementing the second step is $\bigO{N_{\mathcal{N}}+\log d_E}$, where the logarithmic term is the cost of implementing $U^{\Phi}_{E \tilde{E}}$.
Finally, the complexity of the third step is $\bigO{  \upkappa_{\sigma}N_{\sigma}\log\frac{d_E \upkappa_{\mathcal{N}(\sigma)}}{\varepsilon}}$. An alternative for this last step is to consider choosing a threshold $\theta$ higher than $1/\upkappa_{\sigma}$, and approximating the square root function by constant zero below the threshold. Indeed, then choosing $\theta \approx \eps^2/(d_E\upkappa_{\mathcal{N}(\sigma)})$ suffices, resulting in the alternative complexity $\mathcal{O}\Big(  \frac{d_E\upkappa_{\mathcal{N}(\sigma)}}{\eps^2}N_{\sigma}\log\frac{d_E \upkappa_{\mathcal{N}(\sigma)}}{\varepsilon}\Big)$ of the third step. 

{\em Lower bounds}---Our algorithm uses the forward channel unitary $U_{E^{\prime}A\rightarrow EB}^{\mathcal{N}}$ about $\bigO{\sqrt{d_{E}\upkappa_{\mathcal{N}(\sigma)}}}$ times. We now prove that there is no generally applicable algorithm that uses $U_{E^{\prime}A\rightarrow EB}^{\mathcal{N}}$ fewer than $\Omega\Big(d_{E}^{\frac12-\alpha}\upkappa_{\mathcal{N}(\sigma)}^{\alpha}\Big)$ times, for all $\alpha\in [0,\frac12]$, thereby ruling out the possibility of large improvements on our algorithm that would simultaneously improve the dependence on both parameters $d_E$ and $\upkappa_{\mathcal{N}(\sigma)}$. 

We consider solving the problem of unstructured search of $N \geq 2$ elements with only a single marked element. Let $O$ be a search oracle that recognizes the single marked element. Let the input state $\sigma_A$ be the maximally mixed state representing a uniformly random index $i\in [N]$. The forward channel $\mathcal{N}_{A\rightarrow B}$ applies the search oracle and outputs its output, which is equal to $1$ if $i$ is the marked element and is equal to $0$ otherwise. Hence $\mathcal{N}_{A\rightarrow B}(\sigma_{A}) = \textrm{diag}(1-\frac{1}{N},\frac{1}{N})$ and $\upkappa_{\mathcal{N}(\sigma)} = d_E =  N$. Let $\mathcal{P}^{\mathcal{N},\sigma_A}$ be the Petz recovery channel defined from $\mathcal{N}$ and $\sigma_A$ as specified  above. Now applying the exact channel $\mathcal{P}^{\mathcal{N},\sigma_A}$ on the state $\omega_B= \ket{1}\bra{1}$ finds the marked element with certainty. Thus, for every constant $c<1$, applying a $c$-approximate channel $\tilde{\mathcal{P}}^{\mathcal{N},\sigma_A}$ on $\omega_B$ still finds a marked element with probability at least $1-c$. This requires $\Omega(\sqrt{N}) = \Omega\Big(d_{E}^{\frac12-\alpha}\upkappa_{\mathcal{N}(\sigma)}^{\alpha}\Big)$ uses of~$O$, as the well known quantum search lower bound  states \cite{brassard2002AmpAndEst}.


{\em Pretty good measurement}---One can use our algorithm to implement the {\em pretty good measurement} \cite{Belavkin75,Belavkin75a,HolPGM78,Hausladen93bach,hausladen94}, which is a special case of the Petz map. In this application, one is given a set $\{\sigma_{B}^{x}\}_x$ of states  and a probability distribution $p_{X}$. Let $\sigma_{XB}$ denote the following classical--quantum state: $\sigma_{XB}\coloneqq\sum_{x}p_{X}(x)\ketbra{x}{x}_{X}\otimes\sigma_{B}%
^{x}.$ Let $\mathcal{N}_{XB\rightarrow B}\coloneqq\operatorname{Tr}_{X}$ be the partial trace channel that discards system $X$. 

We now plug these choices into \eqref{eq:petz-map}. The adjoint map $(\mathcal{N}_{XB\rightarrow B})^{\dag}$ appends the identity on system $X$. Let $\overline{\sigma}_{B}\coloneqq \mathcal{N}_{XB\rightarrow B}(\sigma_{XB})=\sum_{x}p_{X}(x)\sigma_{B}^{x}$.  The resulting Petz recovery channel is as follows:
\begin{equation*}
\mathcal{P}_{B\rightarrow XB}^{\sigma_{XB},\operatorname{Tr}_{X}}(\omega
_{B})\coloneqq  
\sum_{x} [x]_{X}\otimes p_X(x)\left(  \sigma_{B}^{x}\right)
^{\frac{1}{2}}
\left(\overline{\sigma}_{B}\right)^{-\frac{1}{2}}
\omega_{B}
\left(\overline{\sigma}_{B}\right)^{-\frac{1}{2}}
\left(
\sigma_{B}^{x}\right)  ^{\frac{1}{2}},
\end{equation*}
which is known as the ``pretty good instrument'' \cite{wilde2015recoverabilityInQITheor} and where $[x] \equiv \ketbra{x}{x}$. This is a generalization of the pretty good measurement that has a quantum output in addition to the usual classical measurement output; the PGM is obtained by discarding the quantum output.

We check the necessary assumptions for our technique against what is potentially available for experiments. The isometric extension of the channel $\Tr_X(\cdot)$ is simply the identity. If we have copies of $\sigma_{XB}$ then our algorithm is applicable, but it is more efficient in the case when we can prepare a purification of $\sigma_{XB}$. Applying Theorem \ref{thm:petz}, we arrive at a quantum algorithm implementing the pretty good instrument with performance guarantees as in \eqref{eq:final-diamond-norm-bnd-petz}\ and
\eqref{eq:performance-petz-map}, where%
\begin{equation}
d_{E} =\left\vert X\right\vert ,\quad
\upkappa_{\mathcal{N}(\sigma)}  = \upkappa_{\overline{\sigma}},\quad 
\upkappa_{\sigma}  =\min_{x}p_{X}(x)\upkappa_{\sigma_{B}^{x}}.
\end{equation}

{\em Conclusion}---We have developed a quantum algorithm for implementing the Petz recovery channel and the pretty good measurement. This solves an important open problem in quantum computation, and more generally, it opens up a new research paradigm for realizing fully quantum Bayesian inference on quantum computers.


\begin{acknowledgments}
We gratefully acknowledge the Simons Institute for the Theory of Computing, where part of this work was conducted.  
AG acknowledges funding provided by Samsung Electronics Co., Ltd., for the project ``The Computational Power of Sampling on Quantum
Computers.'' Additional support was provided by the Institute for Quantum Information and Matter, an NSF Physics Frontiers Center (NSF Grant PHY-1733907).
SL was funded by ARO, AFOSR, and IARPA. IM acknowledges support from the  US National Science Foundation under grant number 1910571. 
YQ acknowledges support from a Stanford QFARM fellowship and from an NUS Overseas Graduate Scholarship. MMW acknowledges support from the US National Science Foundation under grant number 1714215, from Stanford QFARM, and from
AFOSR under grant number FA9550-19-1-0369.
\end{acknowledgments}

\bibliography{refs}

\newcommand{\lName}{1}\newcommand{\arxiv}[1]{arXiv:\href{https://arxiv.org/abs/#1}{\ttfamily{#1}}\removefirstdot}\newcommand{\arXiv}[1]{arXiv:\href{https://arxiv.org/abs/#1}{\ttfamily{#1}}\removefirstdot}\def\removefirstdot#1{\if.#1{}\else#1\fi}\providecommand{\multiletter}[1]{#1}\renewcommand{\multiletter}[1]{#1}\DeclareRobustCommand{\dutchPrefix}[2]{#2}\providecommand{\dutchPrefix}[2]{#2}\renewcommand{\dutchPrefix}[2]{#2}\newcommand{\skp}[3]{#2}\newcommand{\focs
  }[1]{\if\lName1\skp{ }{Proceedings of the #1 {IEEE} Symposium on Foundations
  of Computer Science ({FOCS})}{ }\else{FOCS}\fi}\newcommand{\stoc
  }[1]{\if\lName1\skp{ }{Proceedings of the #1 {ACM} Symposium on the Theory of
  Computing ({STOC})}{ }\else{STOC}\fi}\newcommand{\soda }[1]{\if\lName1\skp{
  }{Proceedings of the #1 {ACM-SIAM} Symposium on Discrete Algorithms
  ({SODA})}{ }\else{SODA}\fi}\newcommand{\stacs }[1]{\if\lName1\skp{
  }{Proceedings of the #1 Symposium on Theoretical Aspects of Computer Science
  ({STACS})}{ }\else{STACS}\fi}\newcommand{\itcs }[1]{\if\lName1\skp{
  }{Proceedings of the #1 Innovations in Theoretical Computer Science
  Conference (ITCS)}{ }\else{ITCS}\fi}\newcommand{\fsttcs }[1]{\if\lName1\skp{
  }{Proceedings of the #1 International Conference on Foundations of Software
  Technology and Theoretical Computer Science (FSTTCS)}{
  }\else{FSTTCS}\fi}\newcommand{\ccc }[1]{\if\lName1\skp{ }{Proceedings of the
  #1 {IEEE} Conference on Computational Complexity ({CCC})}{
  }\else{CCC}\fi}\newcommand{\isit }[1]{\if\lName1\skp{ }{Proceedings of the #1
  {IEEE} International Symposium on Information Theory ({ISIT})}{
  }\else{ISIT}\fi}\newcommand{\colt }[1]{\if\lName1\skp{ }{Proceedings of the
  #1 Conference On Learning Theory (COLT)}{ }\else{COLT}\fi}\newcommand{\nips
  }[1]{\if\lName1\skp{ }{Advances in Neural Information Processing Systems #1
  ({NIPS})}{ }\else{NIPS}\fi}\newcommand{\aistats }[1]{\if\lName1\skp{
  }{Proceedings of the #1 International Conference on Artificial Intelligence
  and Statistics ({AISTATS})}{ }\else{AISTATS}\fi}\newcommand{\icml
  }[1]{\if\lName1\skp{ }{Proceedings of the #1 International Conference on
  Machine Learning (ICML)}{ }\else{ICML}\fi}\newcommand{\icalp
  }[1]{\if\lName1\skp{ }{Proceedings of the #1 International Colloquium on
  Automata, Languages, and Programming (ICALP)}{
  }\else{ICALP}\fi}\newcommand{\esa }[1]{\if\lName1\skp{ }{Proceedings of the
  #1 Annual European Symposium on Algorithms (ESA)}{
  }\else{ESA}\fi}\newcommand{\tqc }[1]{\if\lName1\skp{ }{Proceedings of the #1
  Conference on the Theory of Quantum Computation, Communication, and
  Cryptography (TQC)}{}\else{TQC}\fi}\newcommand{\jacm }{\if\lName1\skp{
  }{Journal of the ACM}{ }\else{J. ACM}\fi}\newcommand{\acmta }{\if\lName1\skp{
  }{ACM Transactions on Algorithms}{ }\else{{ACM} Tr.
  Alg}\fi}\newcommand{\acmtct }{\if\lName1\skp{ }{ACM Transactions on
  Computation Theory}{ }\else{ACM Tr. Comp. Th.}\fi}\newcommand{\jams
  }{\if\lName1\skp{ }{Journal of the AMS}{ }\else{J. AMS}\fi}\newcommand{\pams
  }{\if\lName1\skp{ }{Proceedings of the AMS}{ }\else{Proc.
  AMS}\fi}\newcommand{\linalgappl }{\if\lName1\skp{ }{Linear Algebra and its
  Applications}{ }\else{Lin. Alg. \& App.}\fi}\newcommand{\jalgo
  }{\if\lName1\skp{ }{Journal of Algorithms}{ }\else{J.
  Alg.}\fi}\newcommand{\jcss }{\if\lName1\skp{ }{Journal of Computer and System
  Sciences}{ }\else{J. Comp. Sys. Sci.}\fi}\newcommand{\cc }{\if\lName1\skp{
  }{Computational Complexity}{ }\else{Comp. Comp.}\fi}\newcommand{\algor
  }{\if\lName1\skp{ }{Algorithmica}{ }\else{Alg.}\fi}\newcommand{\comb
  }{\if\lName1\skp{ }{Combinatorica}{ }\else{Comb.}\fi}\newcommand{\cacm
  }{\if\lName1\skp{ }{Communications of the ACM}{ }\else{Comm.
  ACM}\fi}\newcommand{\sigart }{\if\lName1\skp{ }{SIGART Bulletin}{
  }\else{SIGART Bull.}\fi}\newcommand{\sigactn }{\if\lName1\skp{ }{SIGACT
  News}{ }\else{SIGACT News}\fi}\newcommand{\eatcsbul }{\if\lName1\skp{
  }{Bulletin of the {EATCS}}{ }\else{Bull. {EATCS}}\fi}\newcommand{\siamrev
  }{\if\lName1\skp{ }{SIAM Review}{ }\else{SIAM Rev.}\fi}\newcommand{\siamjc
  }{\if\lName1\skp{ }{SIAM Journal on Computing}{ }\else{SIAM J.
  Comp.}\fi}\newcommand{\siamjo }{\if\lName1\skp{ }{SIAM Journal on
  Optimization}{ }\else{SIAM J. Opt.}\fi}\newcommand{\siamjdm }{\if\lName1\skp{
  }{SIAM Journal on Discrete Mathematics}{ }\else{SIAM J. Disc.
  Math.}\fi}\newcommand{\siamjnum }{\if\lName1\skp{ }{SIAM Journal on Numerical
  Analysis}{ }\else{SIAM J. Num. Anal.}\fi}\newcommand{\siamjmathanal
  }{\if\lName1\skp{ }{SIAM Journal on Mathematical Analysis}{ }\else{SIAM J.
  Math. Anal.}\fi}\newcommand{\discmath }{\if\lName1\skp{ }{Discrete
  Mathematics}{ }\else{Disc. Math.}\fi}\newcommand{\das }{\if\lName1\skp{
  }{Discrete Applied Mathematics}{ }\else{Disc. App.
  Math.}\fi}\newcommand{\amatstat }{\if\lName1\skp{ }{Annals of Mathematical
  Statistics}{ }\else{Ann. Math. Stat.}\fi}\newcommand{\rms }{\if\lName1\skp{
  }{Russian Mathematical Surveys}{ }\else{Russ. Math.
  Surv.}\fi}\newcommand{\invmath }{\if\lName1\skp{ }{Inventiones Mathematicae}{
  }\else{Inv. Math.}\fi}\newcommand{\jnumber }{\if\lName1\skp{ }{Journal of
  Number Theory}{ }\else{J. Num. Th.}\fi}\newcommand{\toc }{\if\lName1\skp{
  }{Theory of Computing}{ }\else{Th. Comp.}\fi}\newcommand{\cjtcs
  }{\if\lName1\skp{ }{Chicago Journal of Theoretical Computer
  Science}{}\else{Chic. J. Th. Comp. Sci.}\fi}\newcommand{\quantum
  }{\if\lName1\skp{ }{{Quantum}}{ }\else{Quant.}\fi}\newcommand{\cmp
  }{\if\lName1\skp{ }{Communications in Mathematical Physics}{ }\else{Comm.
  Math. Phys.}\fi}\newcommand{\jmp }{\if\lName1\skp{ }{Journal of Mathematical
  Physics}{ }\else{J. Math. Phys.}\fi}\newcommand{\rspa }{\if\lName1\skp{
  }{Proceedings of the Royal Society A}{ }\else{Proc. Roy. Soc.
  A}\fi}\newcommand{\qic }{\if\lName1\skp{ }{Quantum Information and
  Computation}{ }\else{Quant. Inf. \& Comp.}\fi}\newcommand{\physrev
  }{\if\lName1\skp{ }{Physical Review}{ }\else{Phys. Rev.}\fi}\newcommand{\pra
  }{\if\lName1\skp{ }{Physical Review A}{ }\else{Phys. Rev.
  A}\fi}\newcommand{\prb }{\if\lName1\skp{ }{Physical Review B}{ }\else{Phys.
  Rev. B}\fi}\newcommand{\pre }{\if\lName1\skp{ }{Physical Review E}{
  }\else{Phys. Rev. E}\fi}\newcommand{\prx }{\if\lName1\skp{ }{Physical Review
  X}{ }\else{Phys. Rev. X}\fi}\newcommand{\prl }{\if\lName1\skp{ }{Physical
  Review Letters}{ }\else{Phys. Rev. Lett.}\fi}\newcommand{\njp
  }{\if\lName1\skp{ }{New Journal of Physics}{ }\else{New J.
  Phys.}\fi}\newcommand{\prapp }{\if\lName1\skp{ }{Physical Review Applied}{
  }\else{Phys. Rev. Appl.}\fi}\newcommand{\physrep }{\if\lName1\skp{ }{Physics
  Reports}{ }\else{Phys. Rep.}\fi}\newcommand{\rmp }{\if\lName1\skp{ }{Reviews
  of Modern Physics}{ }\else{Rev. Mod. Phys. }\fi}\newcommand{\phystoday
  }{\if\lName1\skp{ }{Physics Today}{ }\else{Phys.
  Today}\fi}\newcommand{\physics }{\if\lName1\skp{ }{Physics}{
  }\else{Phys.}\fi}\newcommand{\nature }{\if\lName1\skp{ }{Nature}{
  }\else{Nat.}\fi}\newcommand{\natcomm }{\if\lName1\skp{ }{Nature
  Communications}{ }\else{Nat. Comm.}\fi}\newcommand{\natphys }{\if\lName1\skp{
  }{Nature Physics}{ }\else{Nat. Phys.}\fi}\newcommand{\npjqi }{\if\lName1\skp{
  }{npj Quantum Information}{ }\else{npj Quant. Inf.}\fi}\newcommand{\scirep
  }{\if\lName1\skp{ }{Scientific Reports}{ }\else{Sci.
  Rep.}\fi}\newcommand{\science }{\if\lName1\skp{ }{Science}{
  }\else{Sci.}\fi}\newcommand{\jpa }{\if\lName1\skp{ }{Journal of Physics A:
  Mathematical and Theoretical}{ }\else{J. Phys. A}\fi}\newcommand{\ijtp
  }{\if\lName1\skp{ }{International Journal of Theoretical Physics}{
  }\else{Int. J. Th. Phys.}\fi}\newcommand{\jmo }{\if\lName1\skp{ }{Journal of
  Modern Optics}{ }\else{J. Mod. Opt.}\fi}\newcommand{\jstatph
  }{\if\lName1\skp{ }{Journal of Statistical Physics}{ }\else{J. Stat.
  Phys.}\fi}\newcommand{\pnas }{\if\lName1\skp{ }{Proceedings of the National
  Academy of Sciences}{ }\else{PNAS}\fi}\newcommand{\lncs }{\if\lName1\skp{
  }{Lecture Notes in Computer Science}{ }\else{L. Notes Comp.
  Sci.}\fi}\newcommand{\lnai }{\if\lName1\skp{ }{Lecture Notes in Artificial
  Intelligence}{ }\else{L. Notes Art. Int.}\fi}\newcommand{\lnm
  }{\if\lName1\skp{ }{Lecture Notes in Mathematics}{ }\else{L. Notes
  Math.}\fi}\newcommand{\tams }{\if\lName1\skp{ }{Transactions of the American
  Mathematical Society}{ }\else{Trans. AMS}\fi}\newcommand{\ieeetit
  }{\if\lName1\skp{ }{{IEEE} Transactions on Information Theory}{ }\else{{IEEE}
  Trans. Inf. Th.}\fi}\newcommand{\iscs }{\if\lName1\skp{ }{International
  Series in Computer Science}{ }\else{Int. Ser. Comp.
  Sci.}\fi}\newcommand{\tocl }{\if\lName1\skp{ }{Theory of Computing Library}{
  }\else{Th. Comp. Lib.}\fi}
\begin{thebibliography}{10}

\bibitem{Belavkin75}
Viacheslav Belavkin.
\newblock Optimal distinction of non-orthogonal quantum signals.
\newblock {\em Radio Engineering and Electronic Physics}, 20:39--47, 1975.

\bibitem{Belavkin75a}
Viacheslav Belavkin.
\newblock Optimal multiple quantum statistical hypothesis testing.
\newblock {\em Stochastics}, 1:315--345, 1975.

\bibitem{HolPGM78}
Alexander~S. Holevo.
\newblock On asymptotically optimal hypothesis testing in quantum statistics.
\newblock {\em Theory of Probability \& Its Applications}, 23(2):411--415,
  1979.

\bibitem{Hausladen93bach}
Paul Hausladen.
\newblock {\em On the Quantum Mechanical Channel Capacity as a Function of the
  Density Matrix}.
\newblock Bachelor's thesis, Williams College, Williamstown, Massachusetts,
  1993.

\bibitem{hausladen94}
Paul Hausladen and William~K. Wootters.
\newblock A `pretty good' measurement for distinguishing quantum states.
\newblock {\em Journal of Modern Optics}, 41(12):2385--2390, 1994.

\bibitem{petz1986sufficientSubalgebrasAndRelativeEntropy}
D\'enes Petz.
\newblock Sufficient subalgebras and the relative entropy of states of a von
  {N}eumann algebra.
\newblock {\em \cmp}, 105(1):123--131, 1986.

\bibitem{petz1988sufficiencyOfChannels}
D\'enes Petz.
\newblock Sufficiency of channels over von {N}eumann algebras.
\newblock {\em Quarterly Journal of Mathematics}, 39(1):97--108, 1988.

\bibitem{OP93}
Masanori Ohya and Denes Petz.
\newblock {\em Quantum Entropy and Its Use}.
\newblock Springer, 1993.

\bibitem{barnum2000reversingQuantumDynamics}
Howard Barnum and Emanuel Knill.
\newblock Reversing quantum dynamics with near-optimal quantum and classical
  fidelity.
\newblock {\em \jmp}, 43(5):2097--2106, 2002.
\newblock \arxiv{quant-ph/0004088}.

\bibitem{hayden2003structureStatesStrongSubaddivityWEquality}
Patrick Hayden, Richard Jozsa, D\'enes Petz, and Andreas Winter.
\newblock Structure of states which satisfy strong subadditivity of quantum
  entropy with equality.
\newblock {\em \cmp}, 246(2):359--374, 2004.
\newblock \arxiv{quant-ph/0304007}.

\bibitem{wilde2017QIT}
Mark~M. Wilde.
\newblock {\em Quantum Information Theory}.
\newblock Cambridge University Press, 2nd edition, 2017.
\newblock \arxiv{1106.1445v8}.

\bibitem{hausladen96}
Paul Hausladen, Richard Jozsa, Benjamin Schumacher, Michael Westmoreland, and
  William~K. Wootters.
\newblock Classical information capacity of a quantum channel.
\newblock {\em Physical Review A}, 54:1869--1876, September 1996.

\bibitem{Hol98}
Alexander~S. Holevo.
\newblock The capacity of the quantum channel with general signal states.
\newblock {\em IEEE Transactions on Information Theory}, 44(1):269--273,
  January 1998.
\newblock \arxiv{quant-ph/9611023}.

\bibitem{SW97}
Benjamin Schumacher and Michael~D. Westmoreland.
\newblock Sending classical information via noisy quantum channels.
\newblock {\em Physical Review A}, 56(1):131--138, July 1997.

\bibitem{BG14}
Salman {Beigi} and Amin {Gohari}.
\newblock Quantum achievability proof via collision relative entropy.
\newblock {\em IEEE Transactions on Information Theory}, 60(12):7980--7986,
  2014.
\newblock \arxiv{1312.3822}.

\bibitem{beigi2015decodingQITviaPetz}
Salman Beigi, Nilanjana Datta, and Felix Leditzky.
\newblock Decoding quantum information via the {P}etz recovery map.
\newblock {\em \jmp}, 57(8):082203, August 2016.
\newblock \arxiv{1504.04449}.

\bibitem{leifer2013TowardsQuantThViaBayesianInference}
Matthew~S. Leifer and Robert~W. Spekkens.
\newblock Towards a formulation of quantum theory as a causally neutral theory
  of {B}ayesian inference.
\newblock {\em \pra}, 88(5):052130, November 2013.
\newblock \arxiv{1107.5849}.

\bibitem{Ng_2010}
Hui~Khoon Ng and Prabha Mandayam.
\newblock Simple approach to approximate quantum error correction based on the
  transpose channel.
\newblock {\em \pra}, 81(6):062342, June 2010.
\newblock \arxiv{0909.0931}.

\bibitem{tyson10}
Jon Tyson.
\newblock Two-sided bounds on minimum-error quantum measurement, on the
  reversibility of quantum dynamics, and on maximum overlap using directional
  iterates.
\newblock {\em Journal of Mathematical Physics}, 51(9):092204, September 2010.
\newblock \arxiv{0907.3386}.

\bibitem{MN2012}
Prabha Mandayam and Hui~Khoon Ng.
\newblock Towards a unified framework for approximate quantum error correction.
\newblock {\em \pra}, 86:012335, July 2012.
\newblock \arxiv{1202.5139}.

\bibitem{DFW15}
Frederic {Dupuis}, Omar {Fawzi}, and Stephanie {Wehner}.
\newblock Entanglement sampling and applications.
\newblock {\em IEEE Transactions on Information Theory}, 61(2):1093--1112,
  2015.
\newblock \arxiv{1305.1316}.

\bibitem{PhysRevA.96.042328}
Joseph~M. Renes.
\newblock Better bounds on optimal measurement and entanglement recovery, with
  applications to uncertainty and monogamy relations.
\newblock {\em Physical Review A}, 96:042328, October 2017.
\newblock \arxiv{1707.01114}.

\bibitem{wilde2015recoverabilityInQITheor}
Mark~M. Wilde.
\newblock Recoverability in quantum information theory.
\newblock {\em \rspa}, 471(2182):20150338, 2015.
\newblock \arxiv{1505.04661}.

\bibitem{Sutter15}
David Sutter, Marco Tomamichel, and Aram~W. Harrow.
\newblock Strengthened monotonicity of relative entropy via pinched {Petz}
  recovery map.
\newblock {\em IEEE Transactions on Information Theory}, 62(5):2907--2913, May
  2016.
\newblock \arXiv{1507.00303}.

\bibitem{Junge15}
Marius Junge, Renato Renner, David Sutter, Mark~M. Wilde, and Andreas Winter.
\newblock Universal recovery from a decrease of quantum relative entropy.
\newblock {\em Annales Henri Poincare}, 19(10):2955--2978, October 2018.
\newblock \arXiv{1509.07127}.

\bibitem{Carlen_2020}
Eric~A. Carlen and Anna Vershynina.
\newblock Recovery map stability for the data processing inequality.
\newblock {\em Journal of Physics A: Mathematical and Theoretical},
  53(3):035204, January 2020.
\newblock \arxiv{1710.02409}.

\bibitem{faulkner2020}
Thomas Faulkner, Stefan Hollands, Brian Swingle, and Yixu Wang.
\newblock Approximate recovery and relative entropy {I}. general von {Neumann}
  subalgebras.
\newblock 2020.
\newblock \arxiv{2006.08002}.

\bibitem{Lami_2018}
Ludovico Lami, Siddhartha Das, and Mark~M. Wilde.
\newblock Approximate reversal of quantum {Gaussian} dynamics.
\newblock {\em Journal of Physics A: Mathematical and Theoretical},
  51(12):125301, February 2018.
\newblock \arxiv{1702.04737}.

\bibitem{swingle2019}
Brian~G. Swingle and Yixu Wang.
\newblock Recovery map for fermionic {Gaussian} channels.
\newblock {\em Journal of Mathematical Physics}, 60(7):072202, July 2019.
\newblock \arxiv{1811.04956}.

\bibitem{CHPSSW19}
Jordan Cotler, Patrick Hayden, Geoffrey Penington, Grant Salton, Brian Swingle,
  and Michael Walter.
\newblock Entanglement wedge reconstruction via universal recovery channels.
\newblock {\em Physical Review X}, 9(3):031011, July 2019.
\newblock \arxiv{1704.05839}.

\bibitem{jia2020}
Hewei~Frederic Jia and Mukund Rangamani.
\newblock Petz reconstruction in random tensor networks.
\newblock June 2020.
\newblock \arxiv{2006.12601}.

\bibitem{PSSY19}
Geoff Penington, Stephen~H. Shenker, Douglas Stanford, and Zhenbin Yang.
\newblock Replica wormholes and the black hole interior, November 2019.
\newblock \arXiv{1911.11977}.

\bibitem{bacon2005optimalMeasurementDiahedralHSP}
Dave Bacon, Andrew~M. Childs, and Wim {\dutchPrefix{Dam}{v}}an~Dam.
\newblock Optimal measurements for the dihedral hidden subgroup problem.
\newblock {\em \cjtcs}, 2006:2, 2006.
\newblock \arxiv{quant-ph/0501044}.

\bibitem{arunachalam2017OptQSampCoplLearn}
Srinivasan Arunachalam and Ronald {\dutchPrefix{Wolf}{d}}e~Wolf.
\newblock Optimal quantum sample complexity of learning algorithms.
\newblock In {\em \ccc{32nd}}, pages 25:1--25:31, 2017.
\newblock \arxiv{1607.00932}.

\bibitem{lloyd2020QPolarDecAlg}
Seth Lloyd, Samuel Bosch, Giacomo De~Palma, Bobak Kiani, Zi-Wen Liu, Milad
  Marvian, Patrick Rebentrost, and David~M. Arvidsson-Shukur.
\newblock Quantum polar decomposition algorithm.
\newblock \arxiv{2006.00841}, 2020.

\bibitem{low2016HamSimQubitization}
Guang~Hao Low and Isaac~L. Chuang.
\newblock Hamiltonian simulation by qubitization.
\newblock {\em \quantum}, 3:163, 2019.
\newblock \arxiv{1610.06546}.

\bibitem{gilyen2018QSingValTransf}
András Gilyén, Yuan Su, Guang~Hao Low, and Nathan Wiebe.
\newblock Quantum singular value transformation and beyond: exponential
  improvements for quantum matrix arithmetics.
\newblock In {\em \stoc{51st}}, pages 193--204, 2019.
\newblock \arxiv{1806.01838}.

\bibitem{lloyd2013QPrincipalCompAnal}
Seth Lloyd, Masoud Mohseni, and Patrick Rebentrost.
\newblock Quantum principal component analysis.
\newblock {\em \natphys}, 10:631--633, 2014.
\newblock \arxiv{1307.0401}.

\bibitem{kimmel2016hamiltonian}
Shelby Kimmel, Cedric Yen-Yu Lin, Guang~Hao Low, Maris Ozols, and Theodore~J.
  Yoder.
\newblock Hamiltonian simulation with optimal sample complexity.
\newblock {\em \npjqi}, 3(1):13, 2017.
\newblock \arxiv{1608.00281}.

\bibitem{cleve2017}
Richard Cleve and Chunhao Wang.
\newblock Efficient quantum algorithms for simulating {Lindblad} evolution.
\newblock In Ioannis Chatzigiannakis, Piotr Indyk, Fabian Kuhn, and Anca
  Muscholl, editors, {\em 44th International Colloquium on Automata, Languages,
  and Programming (ICALP 2017)}, volume~80 of {\em Leibniz International
  Proceedings in Informatics (LIPIcs)}, pages 17:1--17:14, Dagstuhl, Germany,
  2017. Schloss Dagstuhl--Leibniz-Zentrum fuer Informatik.
\newblock \arxiv{1612.09512}.

\bibitem{gilyen2018QSingValTransfThesis}
András Gilyén.
\newblock {\em Quantum Singular Value Transformation \& Its Algorithmic
  Applications}.
\newblock PhD thesis, University of Amsterdam, 2019.

\bibitem{berry2013ExpPrecHamSimSTOC}
Dominic~W. Berry, Andrew~M. Childs, Richard Cleve, Robin Kothari, and
  Rolando~D. Somma.
\newblock Exponential improvement in precision for simulating sparse
  {H}amiltonians.
\newblock In {\em \stoc{46th}}, pages 283--292, 2014.
\newblock \arxiv{1312.1414}.

\bibitem{Kitaev1997}
Alexei~Yu Kitaev.
\newblock {Quantum computations: algorithms and error correction}.
\newblock {\em Russian Mathematical Surveys}, 52(6):1191--1249, December 1997.

\bibitem{apeldoorn2018ImprovedQSDPSolving}
Joran {\dutchPrefix{Apeldoorn}{v}}an~Apeldoorn and Andr{\'a}s Gily{\'e}n.
\newblock Improvements in quantum {SDP}-solving with applications.
\newblock In {\em \icalp{46th}}, pages 99:1--99:15, 2019.
\newblock \arxiv{1804.05058}.

\bibitem{gilyen2018QSingValTransfArXiv}
András Gilyén, Yuan Su, Guang~Hao Low, and Nathan Wiebe.
\newblock Quantum singular value transformation and beyond: exponential
  improvements for quantum matrix arithmetics [full version], 2018.
\newblock \arxiv{1806.01838}.

\bibitem{brassard2002AmpAndEst}
Gilles Brassard, Peter H{\o}yer, Michele Mosca, and Alain Tapp.
\newblock Quantum amplitude amplification and estimation.
\newblock In {\em Quantum Computation and Quantum Information: A Millennium
  Volume}, volume 305 of {\em Contemporary Mathematics Series}, pages 53--74.
  AMS, 2002.
\newblock \arxiv{quant-ph/0005055}.

\end{thebibliography}
\bibliographystyle{unsrt}

\pagebreak



\section*{Supplementary Material}

\section{Introduction}

In this supplementary material, we show how the running time of the algorithm
proposed in the main text can be improved substantially in some cases. That
is, we remove the dependence of the algorithm on the dimension $d_{E}$ of the
environment system $E$, and we replace this parameter with the dimension of
the projection $\Pi_{E}^{\widetilde{\mathcal{N}}(\sigma)}$ onto the support of
the state $\widetilde{\mathcal{N}}_{A\rightarrow E}(\sigma_{A})$, where
$\widetilde{\mathcal{N}}_{A\rightarrow E}$ is the channel complementary to the
original channel $\mathcal{N}_{A\rightarrow B}$ and defined from the unitary
extension $U_{E^{\prime}A\rightarrow EB}^{\mathcal{N}}$. In particular, the complementary channel
is defined by%
\begin{equation}
\widetilde{\mathcal{N}}_{A\rightarrow E}(\omega_{A})=\operatorname{Tr}%
_{B}[U_{E^{\prime}A\rightarrow EB}^{\mathcal{N}}(|0\rangle\!\langle0|_{E^{\prime}}%
\otimes\omega_{A})(U_{E^{\prime}A\rightarrow EB}^{\mathcal{N}})^{\dag}],
\end{equation}
for every input state $\omega_{A}$.

For some examples of $U_{E^{\prime}A\rightarrow EB}^{\mathcal{N}}$ and $\sigma_{A}$,  the dimension of the
subspace onto which $\Pi_{E}^{\widetilde{\mathcal{N}}(\sigma)}$ projects is
substantially smaller than $d_{E}$, i.e.,%
\begin{equation}
\operatorname{Tr}[\Pi_{E}^{\widetilde{\mathcal{N}}(\sigma)}]\ll d_{E},
\end{equation}
and so this replacement can lead to a substantial improvement of the algorithm's
running time in some cases. For example, suppose that $\sigma=|0\rangle\!\langle 0|$ and $\mathcal{N}$ is the qudit erasure channel, defined as
\begin{equation}
\rho \to (1-p) \rho + p |e\rangle\!\langle e|,
\end{equation}
where $p\in [0,1]$ is the erasure probability and $|e\rangle\!\langle e|$ is an erasure symbol (i.e., a rank-one density operator orthogonal to every $d$-dimensional input state $\rho$). The complementary channel in this case is given by\footnote{An isometry for the erasure channel can be implemented as follows: first embed the input $\rho \in \mathbb{C}^{d\times d}$ into $\mathbb{C}^{(d+1)\times (d+1)}$ by extending the input space $\mathbb{C}^{d}$ with $\ket{e}$ resulting in $\tilde{\rho}\in\mathbb{C}^{(d+1)\times (d+1)}$, and append the state $\ketbra{e}{e}$ to it, resulting in $\tilde{\rho}\otimes \ketbra{e}{e}$. Finally, apply the partial SWAP operator $\sum_{j\in[d+1]}\ketbra{jj}{jj}+\sum_{j<k\in[d+1]}\sqrt{1-p}(\ketbra{jk}{jk}+\ketbra{kj}{kj})+i\sqrt{p}(\ketbra{jk}{kj}+\ketbra{kj}{jk})$ to $\tilde{\rho}\otimes \ketbra{e}{e}$.}
\begin{equation}
\rho \to p \rho + (1-p) |e\rangle\!\langle e|,
\end{equation}
and we thus find that $\widetilde{\mathcal{N}}(\sigma) = p |0\rangle \!\langle 0| + (1-p)|e\rangle\!\langle e|$. This density operator has rank two, which is substantially smaller than $ d_E = d+1$, with $d_E$ being the dimension of the environment, i.e., the parameter included in the running time of the quantum algorithm given in the main text.

\textbf{Notation.} Let us remind readers that, as in the main text, we define $\ket{\Gamma}_{E\widetilde{E}} \coloneqq  \sum_{i=1}^{d_E} \ket{i}_{E}\ket{i}_{\tilde{E}}$ which is not a true quantum state vector but is instead only proportional to one. Later we introduce a subspace $F$ of $E$ and use the notation $\ket{\Gamma}_{F\tilde{F}}$ to denote the maximally entangled vector $\sum_{i=1}^{d_F} \ket{i}_{F}\ket{i}_{\tilde{F}}$, where $\tilde{F}\subseteq \tilde{E}$ is some (sub)system isomorphic to~$F$. We also employ the following definition:
\begin{equation}
\tilde{\mathcal{O}}(f(a, b, c))\coloneqq \mathcal{O}(f(a, b, c) \cdot \text {polylog}(f(a, b, c))) .
\label{eq:hide-polylog}
\end{equation}

\section{Overview of the modified algorithm}

Recall from the main text that our algorithm implements an approximation of
the following isometric extension of the Petz recovery channel:%
\begin{equation}\label{eq:Petz_original}
V_{B\rightarrow \tilde{E} A}^{\mathcal{P}}\coloneqq  
\big(\bra{0}_{E^{\prime}}\otimes I_{\tilde{E}}\otimes\sigma_{A}^{\frac{1}{2}}
\big)
(U_{E^{\prime}A\rightarrow EB}^{\mathcal{N}})^{\!\dag}
\big(\ket{\Gamma}_{\!E \tilde{E} }\otimes \left[\mathcal{N}(\sigma_{A})\right]^{-\frac{1}{2}}\big).
\end{equation}
Writing the isometric extension in this way allowed us to break down its
implementation into the following basic steps:

\begin{enumerate}
\item Apply an approximation of $[\mathcal{N}(\sigma_{A})]^{-\frac{1}{2}}$ to
the system $B$ using QSVT.

\item Append a maximally entangled state $\ket{\Phi}_{E\tilde{E}}$ proportional to $\ketbra{\Gamma}{\Gamma}_{E\widetilde{E}}$. (Note that the reduced density operator of $\ket{\Phi}_{E\tilde{E}}$ is proportional to $I_E=\Pi_E$.)

\item Apply $(U_{E^{\prime}A\rightarrow EB}^{\mathcal{N}})^{\dag}$ to systems $EB$.

\item Apply an approximation of $\sigma_{A}^{\frac{1}{2}}$ to the system $A$
using QSVT.

\item The above steps need to be repeated a number of times (detailed in the main text), by performing (robust) oblivious amplitude amplification, in order to boost the success probability of projecting onto $|0\rangle\!\langle0|_{E^{\prime}}$ such that it is nearly equal to one.

\end{enumerate}

This implementation brings the dimension factor $d_{E}$ into the running time due to the usage of the state $\ket{\Phi}_{E\tilde{E}}$, and it is our goal to remove this factor. 
In order to do so, we modify Step~2 of the algorithm: instead of appending a state $\ket{\Phi}_{E\tilde{E}}$ whose reduced density operator on the system $E$ is proportional to $\Pi_E = I_E$, we append a state $\ket{\tPhi}_{E\tilde{E}G}$, such that the reduced density operator of $(I_{E\tilde{E}}\otimes \bra{0}_G) \ket{\tPhi}_{E\tilde{E}G}$ on the system $E$ is proportional to $\Pi_{E}^{\widetilde{\mathcal{N}}(\sigma)}$,
where $\Pi_{E}^{\widetilde{\mathcal{N}}(\kern-0.1mm\sigma\kern-0.15mm)}$ is a projector onto the subspace $  F:=\operatorname{supp}(\tN(\sigma)) \subseteq E$ having dimension $d_F$.

On a high level, the modified algorithm applies an approximation to the following map (cf., Eq.~\eqref{eq:Petz_original})
\begin{equation}
X_{B\rightarrow\widetilde{E}A}^{\mathcal{P}} \coloneqq  
\big(\bra{0}_{E^{\prime}}\otimes I_{\tilde{E}}\otimes\sigma_{A}^{\frac{1}{2}}
\big)
(U_{E^{\prime}A\rightarrow EB}^{\mathcal{N}})^{\!\dag}
\big(
\ket{\Gamma}_{\!F \tilde{F} }
\otimes \left[\mathcal{N}(\sigma_{A})\right]^{-\frac{1}{2}}\big).
\end{equation}
Now we show that $X_{B\rightarrow\widetilde{E}A}^{\mathcal{P}}=V_{B \rightarrow \widetilde{E} A}^{\mathcal{P}}$, proving the correctness of the modified algorithm.
Indeed, without loss of generality we can assume $\ket{\Gamma}_{\!F \tilde{F} }=(\Pi_{E}^{\widetilde{\mathcal{N}}(\sigma)}\otimes I_{\tilde{E}})\ket{\Gamma}_{\!E \tilde{E} }$, and so we have that 
\begin{align}
& V_{B \rightarrow \widetilde{E} A}^{\mathcal{P}}-X_{B \rightarrow \widetilde{E} A}^{\mathcal{P}} \notag \\
&=
\big(\bra{0}_{E^{\prime}}\otimes I_{\tilde{E}}\otimes\sigma_{A}^{\frac{1}{2}}
\big)
(U_{E^{\prime}A\rightarrow EB}^{\mathcal{N}})^{\!\dag} \times \notag \\
&
\qquad \qquad \big(
\big((I_E-\Pi_{E}^{\widetilde{\mathcal{N}}(\sigma)})\otimes I_{\tilde{E}}\big)\ket{\Gamma}_{\!E \tilde{E} }
\otimes \left[\mathcal{N}(\sigma_{A})\right]^{-\frac{1}{2}}\big)\\
&=\big(\underset{M}{\underbrace{(\bra{0}_{E^{\prime}}\otimes\sigma_{A}^{\frac{1}{2}})
(U_{E^{\prime}A\rightarrow EB}^{\mathcal{N}})^{\!\dag}
(I_E-\Pi_{E}^{\widetilde{\mathcal{N}}(\sigma)})}}
\otimes I_{\tilde{E}}\big)\times \notag \\
& \qquad \qquad 
\big(\ket{\Gamma}_{\!E \tilde{E}}\otimes \left[\mathcal{N}(\sigma_{A})\right]^{-\frac{1}{2}}\big).
\end{align}
We conclude by showing that in fact $M=0$ by arguing that $\Tr[M^\dagger M]=0$. 
To this end, observe that 
\begin{equation}
\Tr_B[M^\dagger M]=
(I_E-\Pi_{E}^{\widetilde{\mathcal{N}}(\sigma)})\widetilde{\mathcal{N}}(\sigma_{A})(I_E-\Pi_{E}^{\widetilde{\mathcal{N}}(\sigma)})
=0,
\end{equation}
since  the image of $\Pi_{E}^{\widetilde{\mathcal{N}}(\sigma)}$ is equal to the support of $\widetilde{\mathcal{N}}(\sigma_{A})$, by definition.

\section{Detailed analysis of the modified algorithm}

We now state the modified algorithm more rigorously. To apply the Petz recovery channel $\mathcal{P}^{\mathcal{N},\sigma_A}$ to an input state $\omega_B$, consider the steps given in Algorithm~\ref{algo:Petz_new} below.

\begin{algorithm}[H]
		\textbf{Input:} $U_{A E^{\prime} \rightarrow B E}^{\mathcal{N}}$; $U_{R A}^{\sigma}$, a unitary that prepares a purification of the state $\sigma_{A}$; $U^{\mathcal{N}\left(\sigma_{A}\right)}$, an approximate block-encoding of $\mN(\sigma_A)$.\vspace{5pt}\\
\textbf{Classical pre-processing:} 
\begin{algorithmic}[1]
    \State Compute the circuit $U_{R^{\prime}B}^{\tilde{f}_1(\mathcal{N}(\sigma_A))}$  that enacts a $\left(2 \sqrt{\upkappa_{\mathcal{N}(\sigma)}}, \frac{\tOrd{\eps}}{\sqrt{d_F }}  \right)$-block-encoding of $\left[  \mathcal{N}(\sigma_{A})\right]^{-1/2}$, by applying QSVT to a block-encoding of $ \mathcal{N}(\sigma_{A})$. The relevant block $\tilde{f}_1(\mathcal{N}(\sigma_A))$ satisfies
\begin{equation}
\left \Vert \mathcal{N}(\sigma_{A})^{-\frac{1}{2}} - \tilde{f}_1(\mathcal{N}(\sigma_A)) \right \Vert_{\infty} \leq \frac{\tOrd{\eps}}{\sqrt{d_F}} \, .
\end{equation}
    \State Using Lemma~\ref{lem:proj_actual} below, compute the circuit $U^{\Pi}_{E\tilde{E}G}$
 that prepares a subnormalized state $|\widetilde{\Phi}\rangle_{E\tilde{E}G}$, which approximately purifies $\Pi_{E}^{\widetilde{\mathcal{N}}(\sigma)} \ket{\Gamma}_{E\tilde{E}}$ up to a prefactor of $2\sqrt{\upkappa_{\tN(\sigma)}}$. That is, 
 \begin{equation}
\left \Vert\Pi_{E}^{\widetilde{\mathcal{N}}(\sigma)} \ket{\Gamma}_{E\tilde{E}}  - 2\sqrt{\upkappa_{\tN(\sigma)}}\big(\bra{0}_G\otimes I_{E\tilde{E}}\big) \ket{\tPhi}_{E\tilde{E}G} \right \Vert_2 \leq  \tOrd{\eps}\, .
\end{equation}
  \State  Compute the circuit $U_{R^{\prime \prime} A}^{\tilde{f}_{2}\left(\sigma_{A}\right)}$ that enacts a $\left(2, \frac{\tOrd{\eps}}{\sqrt{d_F \upkappa_{\mN(\sigma)}}} \right)$-block-encoding of $\sigma_A^{-\frac{1}{2}}$, by applying QSVT to a block-encoding of $\sigma_A$. The relevant block $\tilde{f}_{2}\left(\sigma_{A}\right)$ satisfies
  \begin{equation}
\left \Vert  \sigma_A^{1/2} - \tilde{f}_2(\sigma_A) \right \Vert_{\infty}  \leq    \frac{\tOrd{\eps}}{\sqrt{d_F \upkappa_{\mN(\sigma)}}} \, .
\end{equation}
\end{algorithmic}

\textbf{Quantum steps:} 
\begin{algorithmic}[1]
\setcounter{ALG@line}{+3}
\State Perform $O(\sqrt{\upkappa_{\mathcal{N}(\sigma)} \cdot \upkappa_{\widetilde{\mathcal{N}}(\sigma)}})$ rounds of oblivious amplitude amplification on the unitary formed by composing the unitaries prepared in the first three steps
\begin{equation}
U_{R^{\prime \prime} A}^{\tilde{f}_{2}\left(\sigma_{A}\right)} \left(U_{A E^{\prime} \rightarrow B E}^{\mathcal{N}}\right)^{\dagger} \left(U^{\Pi}_{E\tilde{E}G}\otimes U_{R^{\prime}B}^{\tilde{f}_1(\mathcal{N}(\sigma_A))}\right)
\end{equation}
and act with the resultant unitary on the input state of system $B$.
\end{algorithmic}
\textbf{Result:}
The above map acts on an arbitrary input state as an approximation to \eqref{eq:Petz_original}, which is an isometric extension of the Petz recovery channel $\mathcal{P}_{B \rightarrow A}^{\sigma, \mathcal{N}}$.
\caption{\textsf{Petz recovery channel associated with the channel $\mN_{A \to B}$ and the state $\sigma_A$.}
}
\label{algo:Petz_new}
\end{algorithm}

\subsection{Complexity}
\subsubsection{Implementing Step 2}
We now explain how to prepare the state $\ket{\widetilde{\Phi}}_{E\tE G}$ that features in our new Step 2. We require the following definitions and adaptation of a lemma from \cite{apeldoorn2018ImprovedQSDPSolving, gilyen2018QSingValTransfThesis}:

\begin{definition}[Subnormalized density operator] 
A subnormalized density operator $\varrho$ is a positive semi-definite matrix of trace at most one (i.e., $\operatorname{Tr}[\varrho] \leq 1$). In particular, it need not be a valid density operator, which has trace equal to one.
\end{definition}

\begin{definition}[Purification]
A {\em purification} of a subnormalized density operator $\varrho$ is a pure state consisting of three registers, such that tracing out the third register and projecting onto the subspace for which the second register is $|0\rangle$ yields $\varrho .$
\end{definition}

\begin{lemma}[Implementing a purification of a projector]
\label{lem:proj_actual}
Let $q \in (0,1)$, $\Pi$ be a projector, and $\varrho$ a subnormalized density operator acting on the system $E$. Suppose that $q \Pi \preceq \varrho$ and $(I-\Pi) \varrho(I-\Pi)=0$, and suppose that we have access to an a-qubit unitary $U_{\varrho}$ preparing a purification of a subnormalized density operator $\varrho$. Then we can a prepare a pure state $\ket{\widetilde{\Phi}}_{E\tilde{E}G}$ satisfying
\begin{equation}
\left \Vert \big(\bra{0}_G\otimes I_{E\tilde{E}}\big)\ket{\widetilde{\Phi}}_{E\tilde{E}G} - \frac{\sqrt{q}}{2} \big(\Pi_E \otimes U_{\varphi, \tilde{E}}\big)   \ket{\Gamma}_{E\tilde{E}} \right \Vert_2 \leq \eps, 
\label{eq:purify-proj-ineq}
\end{equation}
where $U_{\varphi, \tilde{E}}$ is a unitary acting on the ancillary system $\tilde{E}$ (which is traced over in our applications). We can do so with $\Ord{(1/q)\, \mathrm{polylog}(1/q,1/\eps)}$ queries to $U_{\tilde{\varrho}}$ and its inverse.
\end{lemma}

\begin{proof}
This lemma follows from Lemma~6.4.11 of \cite{gilyen2018QSingValTransfThesis}. From there, we conclude that it is possible to implement a unitary $W_{EG}$ that is a $(1, \widetilde{\mathcal{O}}_{\nu, q}(a), 0)$-block-encoding of an operator~$V_{E}$ satisfying
\begin{equation}
\left\|\left(V-\frac{\sqrt{q}}{2 \sqrt{\varrho}}\right) \Pi\right\|_{\infty} \leq \nu
\label{eq:proof-ineq-key}
\end{equation}
with $\Ord{(1/q)\, \mathrm{polylog}(1/q,1/\eps)}$ queries to $U_{\varrho}$ and its inverse. By the assumptions of the lemma (i.e., $q \Pi \preceq \varrho$ and $(I-\Pi) \varrho(I-\Pi)=0$), the operator $\varrho_E$ is supported on the image of~$\Pi_E$, which implies that $\Pi \varrho \Pi = \varrho$ and $[\Pi,\varrho]=0$.

We now apply the inequality in \eqref{eq:proof-ineq-key} to arrive at the claim in \eqref{eq:purify-proj-ineq}.
Let $\ket{\psi_{\varrho}}_{E\tE}\coloneqq  (\sqrt{\varrho}_E \otimes I_{\tE}) \ket{\Gamma}_{E\tE}$ denote the canonical purification of $\varrho_E$. Then, omitting system labels for conciseness, we find that
\begin{align}
\left\|\left(V-\frac{\sqrt{q}}{2 \sqrt{\varrho}}\right) \Pi\right\|_{\infty} & \leq \nu\\
\Rightarrow \quad
\left \Vert V \Pi  \otimes I  -  \frac{\sqrt{q}}{2\sqrt{\varrho}} \Pi \otimes I  \right \Vert_{\infty} &\leq \nu\\
\Rightarrow \quad \left \Vert   V \otimes I \ket{\psi_{\varrho}} -  \frac{\sqrt{q}}{2\sqrt{\varrho}} \Pi \otimes I  \ket{\psi_{\varrho}} \right \Vert_{2} &\leq \nu\\
\Rightarrow \quad \left \Vert V \otimes I \ket{\psi_{\varrho}} -  \frac{\sqrt{q}}{2} \Pi  \otimes I   \ket{\Gamma} \right \Vert_{2} &\leq \nu\label{eq:last}.
\end{align}
The above steps follow because $\left\|A\right\|_\infty = \left\|A \otimes I\right\|_\infty$, from the definition of the spectral norm, and from the assumption that $\Pi$ commutes with $\varrho$.

Now let $\ket{\varphi_{\varrho}}$ be the actual purification of $\varrho$ prepared by the unitary $U_{\varrho}$, which is related to the canonical purification $\ket{\psi_{\varrho}}$ by a unitary $U_{\varphi}$ acting on the reference system:
\begin{equation}
\ket{\varphi_{\varrho}} =  I \otimes U_{\varphi} \ket{\psi_{\varrho}}.
\end{equation}
This follows by the well known fact that every two purifications of a density operator are related to each other by a  unitary acting on the purifying system \cite{wilde2017QIT}.
Continuing by multiplying each term in \eqref{eq:last} by $ I \otimes U_{\varphi} $ (due to unitary invariance of the norm), we conclude that
\begin{align}\label{eq:approximation}
\left \Vert V \otimes I  \ket{\varphi_{\varrho}} -  \frac{\sqrt{q}}{2} \Pi  \otimes U_{\varphi}  \ket{\Gamma} \right \Vert_2 \leq \nu.
\end{align}

Thus, we can prepare the unnormalized state $(V_E \otimes I_{\tilde{E}}) \ket{\varphi_{\varrho}}_{E\tilde{E}} $ by acting on $\ket{\varphi_{\varrho}}_{E\tilde{E}} \otimes \ket{0}_G$ with $W_{EG}$ and post-selecting on the $\ket{0}_G$ subspace. Therefore the first term in \eqref{eq:approximation} is equal to $\big(\bra{0}_G\otimes I_{E\tilde{E}}\big)\ket{\widetilde{\Phi}}_{E\tilde{E}G}$. The second term in \eqref{eq:approximation} is a purification of the desired projector, which follows because
\begin{equation}
\Tr_{\tilde{E}}[(\Pi_E  \otimes U_{\varphi,\tilde{E}}) \ketbra{\Gamma}{\Gamma}_{E\tilde{E}} (\Pi_E \otimes (U_{\varphi,\tilde{E}})^{\dagger})]    = \Pi_E.
\end{equation}
(We note that $\Pi$ projects onto some subspace of the system $E$, not necessarily the whole system.) Thus, \eqref{eq:approximation} implies that we can prepare a state $\ket{\widetilde{\Phi}}_{E\tilde{E}G}$ which, after postselection on $\ket{0}_G$, well-approximates  the desired purification $(\Pi\otimes I ) \ket{\Gamma}$ up to a scale factor and a unitary $U_{\varphi,\tilde{E}}$ acting on a reference system. Furthermore, the unitary
\begin{equation}
U^{\Pi} \coloneqq  W_{EG} U_{\varrho,E\tE},
\end{equation}
acting on the initial state $\ket{0}_{E\tE G}$, prepares this state, where $U_{\varrho,E\tE}$ is the unitary defined in the lemma statement.
\end{proof}

\bigskip

We now explain how to use Lemma~\ref{lem:proj_actual} to implement Step 2. First, we need to slightly modify our access assumptions: instead of our previous assumption of access to the block-encoding $U^{\sigma_A}$ directly, we assume we have access to  a unitary $U^{\sigma}_{RA}$ that prepares a purification of the state $\sigma_A$. (As mentioned in the main text, with two uses of $U^{\sigma}_{RA}$, we can prepare $U^{\sigma_A}$.) The unitary $U^{\sigma}_{RA}$ takes the role of the unitary $U_{\varrho}$ in Lemma~\ref{lem:proj_actual}. Accordingly, we keep track of the number of uses of $U^{\sigma}_{RA}$ and denote its gate complexity as $N_{\sigma}'$.\footnote{Here the prime $'$ reflects that it is closely related to, and at most twice as large as, $N_{\sigma}$.}

Now, it is easy to see that the unitary
\begin{equation}\label{eq:u_proj}
U_{\tilde{\rho}} = U^{\mathcal{N}}_{A\rightarrow BE} \circ U^{\sigma}_{RA}
\end{equation}
acting on $\ketbra{0}{0}$
prepares a state $\ketbra{\psi}{\psi}_{RBE}$ satisfying
\begin{equation}
\Tr_{RB}(\ket{\psi}\!\bra{\psi}_{RBE}) = \widetilde{\mathcal{N}}_{A\rightarrow E}(\sigma_A) =: \varrho_E.
\end{equation}
Thus, in Lemma~\ref{lem:proj_actual}, we set 
\begin{equation}
\varrho \leftarrow \varrho_E.   
\end{equation}
Our goal is to implement a unitary that prepares a block-encoded projector onto $\operatorname{supp}(\varrho_E)$, and so we also set 
\begin{equation}
\Pi \leftarrow \Pi^{\widetilde{\mathcal{N}}{(\sigma)}}_E.
\end{equation}
That is, considering \eqref{eq:approximation}, the state we can prepare is close to a purification of the  projector~$\Pi^{\widetilde{\mathcal{N}}{(\sigma)}}_E$. Also, by the condition of Lemma~\ref{lem:proj_actual} that $q\Pi\preceq \varrho$, we see that it also suffices to set
\begin{equation}
q \leftarrow \lambda_{\rm min}(\widetilde{\mathcal{N}}(\sigma)).
\end{equation}
Putting together Lemma~\ref{lem:proj_actual} with  \eqref{eq:u_proj}, we see that with a gate complexity of \begin{equation}
\Ord{\upkappa_{\widetilde{\mathcal{N}}(\sigma)} (N_{\mathcal{N}} + N_{\sigma}') \text{polylog}(\upkappa_{\widetilde{\mathcal{N}}(\sigma)}, 1/\eps)},
\end{equation}
we can prepare the desired state $\ket{\widetilde{\Phi}}_{E\tilde{E}G}$.

\subsubsection{Detailed Complexity Analysis}

We now analyze the complexity of the entire modified algorithm:

\begin{itemize}

    \item Step 1 has gate complexity $$\bigO{  \upkappa_{\mathcal{N}(\sigma)}N_{\mathcal{N}(\sigma)}\mathrm{polylog}\left(d_F, \upkappa_{\mathcal{N}(\sigma)}, 1/\eps \right)}.$$ 

\item Step 2 has gate complexity $$\mathcal{O}\left(\upkappa_{\widetilde{\mathcal{N}}(\sigma)} \mathrm{polylog}\left(\upkappa_{\widetilde{\mathcal{N}}(\sigma)},  1/\eps \right)(N_{\mathcal{N}} + N_{\sigma}')\right).$$ The gate complexity to implement $\left(U_{A E^{\prime} \rightarrow B E}^{\mathcal{N}}\right)^{\dagger}$ is, by definition, $N_{\mathcal{N}}$. 

\item Step 3 has gate complexity $$\mathcal{O}\!\left(\upkappa_{\sigma} N_{\sigma}' \mathrm{polylog}\left(d_F,  \upkappa_{\mathcal{N}(\sigma)}, 1/\eps \right)\right).$$ 
\end{itemize}

Finally, for the actual quantum computational  implementation, oblivious amplitude amplification multiplies the overall gate complexity by the number of rounds of amplification $\Ord{\sqrt{\upkappa_{\mathcal{N}(\sigma)} \upkappa_{\widetilde{\mathcal{N}}(\sigma)}}} $ (which is just the product of the accumulated subnormalization factors). The overall gate complexity is thus
\begin{equation}
\widetilde{\mathcal{O}}\left(\sqrt{\upkappa_{\mathcal{N}(\sigma)} \upkappa_{\widetilde{\mathcal{N}}(\sigma)}} \left(\upkappa_{\mathcal{N}(\sigma)}N_{\mathcal{N}(\sigma)}  + (\upkappa_{\sigma} + \upkappa_{\widetilde{\mathcal{N}}(\sigma)} )N_{\sigma}' + \upkappa_{\widetilde{\mathcal{N}}(\sigma)} N_{\mathcal{N}} \right)\right).
\end{equation}
where the symbol $\,\widetilde{}\,$ hides $\text{polylog}(d_F, \upkappa_{\widetilde{\mathcal{N}}(\sigma)}, \upkappa_{\mathcal{N}(\sigma)}, 1/\eps)$ factors, as indicated in \eqref{eq:hide-polylog}.

The error analysis proceeds similarly to that in the main text. The overall unitary resulting from Steps 1, 2, and 3 is as follows:
\begin{multline}\label{eq:block}
U^{\tilde{f}_2(\sigma_A)}_{R''A} (U^{\mathcal{N}}_{E'A \rightarrow EB})^{\dagger} U^{\widetilde{\Pi}}_{F\tilde{F}G} \otimes U^{\tilde{f} _1(\mathcal{N}(\sigma_A))}_{R'B} = \\
\left[\begin{array}{cc}\frac{1}{4 \sqrt{\upkappa_{\mathcal{N}(\sigma)}} \cdot \, 2 \sqrt{\upkappa_{\widetilde{\mathcal{N}}(\sigma)}}}\tilde{X}^{\mathcal{P}}_{B\rightarrow \tilde{E}A} & \cdot \\
\cdot & \cdot
\end{array}\right]
\end{multline}
where the relevant block is the $\ket{0}_{E'G}$ subspace and
\begin{multline}
\tilde{X}^{\mathcal{P}}_{B\rightarrow \tilde{E}A} \coloneqq 
 \tilde{f}_2(\sigma_A) \left(U_{E'A \rightarrow E B}^{\mathcal{N}}\right)^{\dagger} \tilde{f}_1(\mathcal{N}(\sigma_A)) 2\sqrt{\upkappa_{\tN(\sigma)}}\times \\
\big(\bra{0}_G\otimes I_{E\tilde{E}}\big)\ket{\widetilde{\Phi}}_{E \tE G} \otimes I_B
\end{multline}
(cf., Eqs. (15) and (16) in the main text). The oblivious amplitude amplification in Step 4 boosts the probability of success, which is currently $\tOrd{\sqrt{\upkappa_{\mathcal{N}(\sigma)}}  \sqrt{\upkappa_{\widetilde{\mathcal{N}}(\sigma)}}}$, to $\approx 1$, thus implementing the map $\tilde{X}_{B\rightarrow\widetilde{E}A}^{\mathcal{P}} $ with the boosted probability. However, this is an approximation of the ideal map (cf., Eq.~\eqref{eq:Petz_original}) given by
\begin{equation}\label{eq:ideal_map}
X_{B\rightarrow\widetilde{E}A}^{\mathcal{P}} = \sigma_{A}^{\frac{1}{2}%
}(U^{\mathcal{N}}_{E^{\prime}A\rightarrow EB})^{\dag}[\mathcal{N}(\sigma_{A})]^{-\frac{1}%
{2}}\ket{\Gamma}_{F\widetilde{F}}\otimes I_{B}.
\end{equation}
(Note that any map which differs from \eqref{eq:ideal_map} by a unitary on the $\tilde{F}$ system works just as well; hence, without loss of generality, we have set this unitary to the identity.) 

We now compute the $\infty$-norm distance between these two operators,  and we omit the identity on system $B$ to save space:
\begin{widetext}
\begin{align}
& \left \Vert X_{B\rightarrow\widetilde{E}A}^{\mathcal{P}}- \tilde{X}^{\mathcal{P}}_{B\rightarrow\widetilde{E}A}  \right \Vert_{\infty} \notag \\
& = \left \Vert  \sigma_{A}^{\frac{1}{2}%
}(U^{\mathcal{N}}_{E^{\prime}A\rightarrow EB})^{\dag}[\mathcal{N}(\sigma_{A})]^{-\frac{1}%
{2}}|\Gamma\rangle_{F\widetilde{F}%
} - \tilde{f}_2(\sigma_A) \left(U_{E'A \rightarrow E B}^{\mathcal{N}}\right)^{\dagger} \tilde{f}_1(\mathcal{N}(\sigma_A)) 2\sqrt{\upkappa_{\tN(\sigma)}} \big(\bra{0}_G\otimes I_{E\tilde{E}}\big) \ket{\tPhi}_{F\tilde{F}G}
\right \Vert_{\infty}\\
& \leq \left \Vert  \sigma_A^{1/2} - \tilde{f}_2(\sigma_A) \right \Vert_{\infty} \left \Vert (U^{\mathcal{N}}_{E^{\prime}A\rightarrow EB})^{\dag}[\mathcal{N}(\sigma_{A})]^{-\frac{1}%
{2}}(|\Gamma\rangle_{F\widetilde{F}%
})\right \Vert_2 \notag \\
&\qquad + \left \Vert \tilde{f}_2(\sigma_A)
\left(U_{E'A \rightarrow E B}^{\mathcal{N}}\right)^{\dagger}  \right \Vert_{\infty} \left \Vert [\mathcal{N}(\sigma_{A})]^{-\frac{1}{2}} - \tilde{f}_1(\mathcal{N}(\sigma_A)) \right \Vert_{\infty} \left \Vert \ket{\Gamma}_{F\tilde{F}} \right \Vert_2\notag \\
&\qquad + \left \Vert  \tilde{f}_2(\sigma_A) \left(U_{E'A \rightarrow E B}^{\mathcal{N}}\right)^{\dagger} \tilde{f}_1(\mathcal{N}(\sigma_A)) \right \Vert_{\infty} \left \Vert \ket{\Gamma}_{F\tilde{F}}  - 2\sqrt{\upkappa_{\tN(\sigma)}}\big(\bra{0}_G\otimes I_{E\tilde{E}}\big) \ket{\tPhi}_{F\tilde{F}G} \right \Vert_2
\end{align}
\end{widetext}
Then, noting that 
\begin{align}
\left \Vert (U^{\mathcal{N}}_{E^{\prime}A\rightarrow EB})^{\dag}[\mathcal{N}(\sigma_{A})]^{-\frac{1}%
{2}}(|\Gamma\rangle_{F\widetilde{F}%
})\right \Vert_2 &\leq \sqrt{d_F \upkappa_{\mathcal{N}(\sigma)}} ,\\
\left \Vert \tilde{f}_2(\sigma_A)
\left(U_{E'A \rightarrow E B}^{\mathcal{N}}\right)^{\dagger}  \right \Vert_{\infty} \left \Vert \ket{\Gamma}_{F\tilde{F}} \right \Vert_2 &\leq 2 \sqrt{d_F}, \\
\left \Vert  \tilde{f}_2(\sigma_A) \left(U_{E'A \rightarrow E B}^{\mathcal{N}}\right)^{\dagger} \tilde{f}_1(\mathcal{N}(\sigma_A)) \right \Vert_{\infty} \leq 4 ,
\end{align}
we see that if we choose
\begin{equation}
\left \Vert \mathcal{N}(\sigma_{A})^{-\frac{1}{2}} - \tilde{f}_1(\mathcal{N}(\sigma_A)) \right \Vert_{\infty} \leq \frac{\tOrd{\eps}}{\sqrt{d_F}}  \quad \text{(Step 1 error)},
\end{equation}
\begin{multline}
\left \Vert \ket{\Gamma}_{F\tilde{F}}  - 2\sqrt{\upkappa_{\tN(\sigma)}}\big(\bra{0}_G\otimes I_{E\tilde{E}}\big) \ket{\tPhi}_{F\tilde{F}G} \right \Vert_2 
\leq \tOrd{\eps} \\
\quad \text{(Step 2 error)},
\end{multline}
\begin{equation}
\left \Vert  \sigma_A^{1/2} - \tilde{f}_2(\sigma_A) \right \Vert_{\infty}  \leq    \frac{\tOrd{\eps}}{\sqrt{d_F \upkappa_{\mN(\sigma)}}} \quad \text{(Step 3 error)},
\end{equation}
then we obtain a final approximation error of 
\begin{equation}
\left \Vert X_{B\rightarrow\widetilde{E}A}^{\mathcal{P}}- \tilde{X}^{\mathcal{P}}_{B\rightarrow\widetilde{E}A}  \right \Vert_{\infty} \leq \tOrd{\eps},
\end{equation}
which means the actual error in the relevant block of Eq.~\eqref{eq:block} is $\frac{\eps}{\sqrt{\upkappa_{\mN(\sigma)}\upkappa_{\tN(\sigma)}}}$. The $\sqrt{\upkappa_{\mN(\sigma)}\upkappa_{\tN(\sigma)}}$ rounds of oblivious amplitude amplification then magnify the overall error to $\tOrd{\eps}$. 

\end{document}